\documentclass[twocolumn,amsmath,raggedbottom]{aastex702}
\usepackage{needspace,etoolbox}

\newcommand{\qq}{2026~QQ$_8$}
\newcommand{\cdthird}{2020~CD$_3$}
\newcommand{\ptfive}{2024~PT$_5$}

\shorttitle{VLT, Keck, and Gemini characterization of 2026 QQ8}
\shortauthors{Bolin et al.}
\journalinfo{2026 November 22}

\defcitealias{MPEC2026V185}{\mbox{2026-V185}}
\defcitealias{MPEC2026V273}{\mbox{2026-V273}}
\defcitealias{MPEC2026V323}{\mbox{2026-V323}}
\defcitealias{MPEC2026V337}{\mbox{2026-V337}}
\defcitealias{MPEC2026W10}{\mbox{2026-W10}}
\defcitealias{MPEC2026W64}{\mbox{2026-W64}}
\begin{document}
\title{VLT, Keck, and Gemini characterization of the temporarily captured flyby \qq}
\correspondingauthor{Bryce T. Bolin}
\author[0000-0002-4950-6323]{Bryce T. Bolin}
\affiliation{Eureka Scientific, Oakland, CA 94602, USA}
\email[show]{bolin.astro@gmail.com}
\author[0000-0002-2250-8687]{Leo Y. Alcorn}
\affiliation{W. M. Keck Observatory, Waimea, HI 96743, USA}
\email{lalcorn@keck.hawaii.edu}
\author[0000-0003-0773-1888]{David J. Tholen}
\affiliation{Institute for Astronomy, University of Hawai{\textquoteleft}i at M\=anoa, 2680 Woodlawn Drive, Honolulu, HI 96822, USA}
\email{tholen@ifa.hawaii.edu}
\author[0000-0002-1428-7036]{Brian C. Lemaux}
\affiliation{International Gemini Observatory/NSF NOIRLab, Hilo, HI 96720, USA}
\email{brian.lemaux@noirlab.edu}
\author[0000-0002-3370-187X]{Eunchong Kim}
\affiliation{International Gemini Observatory/NSF NOIRLab, Hilo, HI 96720, USA}
\email{eunchong.kim@noirlab.edu}
\author[0009-0000-4044-4640]{Leila Alamos}
\affiliation{International Gemini Observatory/NSF NOIRLab, Hilo, HI 96720, USA}
\email{leila.alamos@noirlab.edu}
\author[0000-0003-2287-158X]{Jennie L. Berghuis}
\affiliation{International Gemini Observatory/NSF NOIRLab, Hilo, HI 96720, USA}
\email{jennie.berghuis@noirlab.edu}
\author[0000-0002-9486-4840]{Henri M.J. Boffin}
\affiliation{European Southern Observatory, Garching bei M\"unchen, Germany}
\email{hboffin@eso.org}
\author[0000-0002-1857-2088]{Boris Häußler}
\affiliation{European Southern Observatory, Alonso de Cordova 3107, Casilla 19001, Santiago, Chile}
\email{Boris.Haeussler@eso.org}
\author[0000-0002-1823-3975]{Anni Kasikov}
\affiliation{European Southern Observatory, Alonso de Cordova 3107, Casilla 19001, Santiago, Chile}
\email{Anni.Kasikov@eso.org}
\author[0009-0006-0758-2113]{Jen Miller}
\affiliation{International Gemini Observatory/NSF NOIRLab, Hilo, HI 96720, USA}
\email{jen.miller@noirlab.edu}
\author[0000-0001-6540-0767]{Thallis Pessi}
\affiliation{European Southern Observatory, Alonso de Cordova 3107, Casilla 19001, Santiago, Chile}
\email{Thallis.Pessi@eso.org}
\author[0009-0001-9216-826X]{Mat\'ias Rodr\'iguez-Henr\'iquez}
\affiliation{International Gemini Observatory/NSF NOIRLab, Hilo, HI 96720, USA}
\email{matias.rodriguez@noirlab.edu}
\author[0000-0002-7444-5315]{Elyar Sedaghati}
\affiliation{European Southern Observatory, Alonso de Cordova 3107, Casilla 19001, Santiago, Chile}
\email{esedagha@eso.org}
\author[0000-0002-4434-2307]{Andrew W. Stephens}
\affiliation{International Gemini Observatory/NSF NOIRLab, Hilo, HI 96720, USA}
\email{andrew.stephens@noirlab.edu}
\author[0009-0003-1592-5102]{Matt Wahl}
\affiliation{W. M. Keck Observatory, Waimea, HI 96743, USA}
\email{mwahl@keck.hawaii.edu}

\begin{abstract}
Temporarily captured flybys (TCFs) enter Earth's Hill radius while bound to the Earth--Moon system and escape before completing one revolution in Sun--Earth corotating geocentric longitude, whereas temporarily captured orbiters complete at least one revolution. Very Large Telescope, Keck~I, and Gemini North photometry of \qq\ (November~7--21) gives Sloan Digital Sky Survey--equivalent colors $g-r=0.594^{+0.056}_{-0.053}$, $g-i=0.840^{+0.066}_{-0.065}$, and $i-z=-0.104^{+0.067}_{-0.057}$~mag. Apollo~17 basalt slabs 75075,197 and 71055,239 closely match these colors ($\chi^2_\nu=0.57$ and $0.89$, respectively). The Bus--DeMeo Sq- and K-type templates give $\chi^2=2.61$ and $2.70$ for three degrees of freedom ($\chi^2_\nu=0.87$ and $0.90$, respectively). With revised Keck timing uncertainties, 71 positions (August~19--November~21) give an outward acceleration of $(1.363\pm0.090)\times10^{-10}$~au~day$^{-2}$ at 1~au, implying an area-to-mass ratio of $(6.02\pm0.40)\times10^{-4}$~m$^2$~kg$^{-1}$ for a unit radiation-pressure coefficient. Osculating heliocentric elements (epoch 2026 September~20) are $a=1.015$~au, $e=0.023$, $i=0.298^\circ$. For pre-encounter analogues, NEOMOD3 assigns source probabilities of 78.8\% to the $\nu_6$ resonance and 20.6\% to Jupiter's 3:1 resonance. Assuming a spherical shape and $H=31.37\pm0.34$, the diameter, density, and mass are ($1.44^{+0.28}_{-0.23}\,\mathrm{m}$, $1.73^{+0.36}_{-0.30}\,\mathrm{g\,cm^{-3}}$, ${2.73^{+1.19}_{-0.82}\times10^3\,\mathrm{kg}}$) for the Sq albedo ($p_V=0.243\pm0.039$), ($1.62^{+0.28}_{-0.24}\,\mathrm{m}$, $1.54^{+0.29}_{-0.24}\,\mathrm{g\,cm^{-3}}$, ${3.43^{+1.30}_{-0.94}\times10^3\,\mathrm{kg}}$) for NEOMOD3 ($p_V=0.191$), and ($2.49^{+0.42}_{-0.36}\,\mathrm{m}$, $1.00^{+0.19}_{-0.16}\,\mathrm{g\,cm^{-3}}$, ${8.09^{+3.06}_{-2.22}\times10^3\,\mathrm{kg}}$) for lunar slabs ($p_V=0.081$). With radiation pressure, nominal capture spans 224.96~days (April~26--December~7). Corotating longitude advances 0.76 turns, classifying \qq\ as a TCF, while the three-dimensional corotating path tangent turns 1.33 revolutions. After 10,000~yr, $\sim$800 of 1,000 clones survive in each model, with $\sim$200 colliding with Earth. Such captures let us test lunar and asteroidal origins for meter-scale Earth co-orbitals.
\end{abstract}

\section{Introduction}\label{sec:intro}

Kriszti\'an S\'arneczky discovered \qq\ on 2026 August~19 with the 0.60~m Schmidt telescope at Piszk\'estet\H{o}, Hungary (station K88) \citep{MPEC2026U95}. The Minor Planet Center (MPC) listed it as Sar2901 on the Near-Earth Object (NEO) Confirmation Page. Because the initial orbit allowed an artificial origin, the MPC withdrew the listing on September~2 but kept the astrometry public and requested follow-up under Sar2901 (Minor Planet Electronic Circular [MPEC] 2026-R09) \citep{MPEC2026R09}.

David J. Tholen recovered Sar2901 on October~9 at 14:08:38 Coordinated Universal Time (UTC) with the University of Hawai{\textquoteleft}i 88-inch (2.24~m) telescope on Maunakea (T12). The recovery and September~5 and 19 positions from the Lowell Discovery Telescope extended the arc to 51~days. The MPC credited these observations with establishing a natural origin \citep{MPC2026October}. On October~29 the MPC designated the object \qq\ and announced its temporary capture by Earth. Its solar radiation pressure (SRP) fit gave an effective area-to-mass ratio of $(7.94\pm3.41)\times10^{-4}$~m$^2$~kg$^{-1}$ ($1\sigma$), consistent with compact rock \citep[MPEC~2026-U95]{MPEC2026U95}.

We define temporary capture by negative geocentric two-body energy within three Earth Hill radii \citep{Granvik2012,Granvik2013,Fedorets2017}. In Sun--Earth corotating geocentric longitude, a temporarily captured orbiter (TCO) completes at least one revolution. A temporarily captured flyby (TCF) enters one Hill radius while bound and escapes before completing a revolution \citep{Fedorets2017}. A TCO that enters one Hill radius is a minimoon \citep{Jedicke2025}.

Temporary capture brings meter-scale bodies within reach of ground-based telescopes \citep{Granvik2012,Granvik2013,Fedorets2017,Jedicke2018}, although rapid sky motion and short visibility windows limit recovery \citep{Bolin2014,Fedorets2020survey}. Lunar fragments can enter temporary capture upon ejection or after orbiting the Sun. Predicted capture counts depend on ejecta sizes and speeds \citep{Jedicke2025}. Some ejecta enter Earth's 1:1 resonance \citep{Sfair2025}, where models give them lower typical eccentricities and inclinations than main-belt co-orbitals \citep{AlessiJedicke2026}.

For \cdthird, \citet{Bolin2020CD3} measured a rise in reflectance from 0.43 to $0.76\,\mu$m and a decline near $0.9\,\mu$m with the Low Resolution Imaging Spectrometer (LRIS) on Keck~I. Both V-type asteroids and some lunar rocks show this pattern. \citet{Fedorets2020CD3} used broadband colors and the astrometric area-to-mass estimate to establish its natural origin. Archival astrometry traced its capture to a close lunar encounter on 2017 September~15 \citep{Naidu2021}.

During its 2024 September--November encounter, \ptfive\ met the three-Hill-radius flyby criterion but remained outside one Hill radius \citep{delaFuenteMarcos2024PT5,delaFuenteMarcos2025PT5}. Gemini North optical colors favored lunar rocks over S-complex asteroids \citep{Bolin2025PT5}, and the visible and near-infrared spectrum showed pyroxene bands consistent with lunar material \citep{Kareta2025}. In 2025 January, Seimei optical colors overlapped both groups, while near-infrared photometry with the Keck~I Multi-Object Spectrometer for Infrared Exploration (MOSFIRE) favored lunar rocks \citep{Beniyama2025}.

We use Very Large Telescope (VLT), Keck~I, and Gemini North colors to test whether \qq\ resembles exposed lunar basalt or an asteroid. The astrometry constrains its capture duration and radial acceleration. We combine the acceleration with albedos for the candidate surfaces to infer bulk density.

\section{Observations}\label{sec:photometry}

\begin{figure*}[t]
\centering
\includegraphics[width=\textwidth]{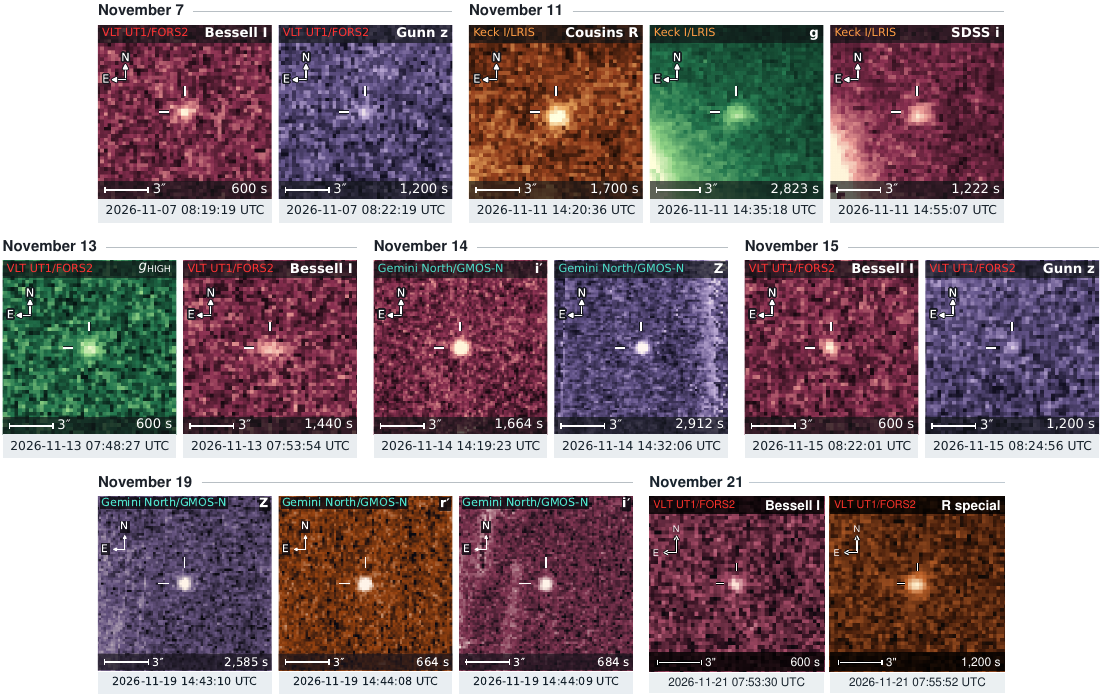}
\caption{Target-aligned stacks of \qq\ in chronological order. Integration time appears inside each stamp and exposure-weighted UTC midpoint below. The FORS2 stacks contain ten frames per filter on November~7, 15, and 21, and five $g$ and six $I$ frames on November~13 after excluding $g$ exposure~7, crossed by a stellar trail. The November~21 stacks retain trail-subtracted $I$ exposure~17 and the final, unbracketed $R$ exposure. Table~\ref{tab:obslog} includes the November~14 FORS2 $g$ and $R$ sequences. LRIS stacks on November~11 contain 16 Cousins $R$, 23 $g$, and nine SDSS $i$ frames. GMOS-N stacks contain four $i'$ and seven $Z$ frames on November~14, and eleven $Z$, four $r'$, and four $i'$ on November~19, excluding trail-crossed $Z$ frames N20261114S0078 and N20261119S0164. Ticks mark the November~7 centroid ($0\farcs38$ west and $0\farcs13$ north of the prediction) and the predicted position on other nights. Each $12\arcsec\times12\arcsec$ stamp has north up, east left, a $3\arcsec$ scale bar, and an inverse-hyperbolic-sine stretch scaled to the local sky. Native scales are $0\farcs252$, $0\farcs270$, and $0\farcs1614$ per pixel for FORS2, LRIS, and GMOS-N. Tints identify filters. Red, orange, and turquoise labels mark VLT, Keck, and Gemini, respectively.}\label{fig:imaging}
\end{figure*}

We observed \qq\ on seven nights from 2026 November~7 to 21 UTC (Table~\ref{tab:obslog}). At VLT Unit Telescope~1 (UT1) we used the Focal Reducer and Low Dispersion Spectrograph~2 (FORS2) \citep{Appenzeller1998FORS} under European Southern Observatory (ESO) program 117.2C7T.001 (principal investigator [PI] B.~T.~Bolin). We bracketed exposures in alternating filters as for 2025~SC$_{79}$ \citep{Bolin2026SC79}, pairing Bessell $I$ and Gunn $z$ on November~7 and 15, $g_{\rm HIGH}$ and $I$ on November~13, $g_{\rm HIGH}$ and $R_{\rm SPECIAL}$ on November~14, and $I$ and $R_{\rm SPECIAL}$ on November~21 \citep{ESOFORS2Filters}.

On November~11 the Keck~I/LRIS blue arm recorded $g$ while the red arm recorded Cousins $R_{\rm C}$ or Sloan Digital Sky Survey (SDSS) $i$ (target-of-opportunity program 2026B\_K450, PI L.~Y.~Alcorn) \citep{Oke1995LRIS,McCarthy1998LRISBlue}. We used the same two-arm technique for 1998~KY$_{26}$ and Neptunian Trojans \citep{Bolin2025KY26,Bolin2023Trojans}. With the Gemini Multi-Object Spectrograph on Gemini North (GMOS-N) \citep{Hook2004GMOS}, we obtained $i'$ and $Z$ on November~14 and interleaved $r'$, $i'$, and $Z$ on November~19 (Director's Discretionary program GN-2026B-DD-108, PI B.~T.~Bolin). We used GMOS at Gemini South to measure $i-Z$ for 2024~YR$_4$ \citep{Bolin2025YR4}.

Our Python routines calibrated each FORS2 chip separately following \citet{Boffin2025FORS2Imaging}. We corrected the prescan level, trimmed the image, subtracted the master bias, and divided by the normalized sky flat for that chip and filter (Appendix~\ref{app:fors2cal}).

We measured fluxes and positions in individual exposures and stacked along the predicted motion (Figure~\ref{fig:imaging}), using all 20 frames on November~21. For FORS2 and LRIS, we fitted seeing profiles convolved with the residual target motion and checked for foreground contamination by subtracting same-filter templates built from the other exposures. We calibrated GMOS-N against trailed field stars and measured \qq\ by aperture photometry on November~14 and profile fitting on November~19 (Appendix~\ref{app:photred}).

\section{Observational results}\label{sec:colors}

\subsection{Time-matched colors}

Table~\ref{tab:nightphot} gives nightly FORS2 and GMOS-N colors on the Panoramic Survey Telescope and Rapid Response System~1 (Pan-STARRS1, or PS1) absolute spectrophotometric (AB) magnitude system \citep{OkeGunn1983AB,Tonry2012}. LRIS colors use SDSS-equivalent $g,i$ (AB) and Cousins $R_{\rm C}$ (Vega). Color intervals span the 16th to 84th percentiles.

On November~13 a stellar trail crossed the target aperture in FORS2 $g$ exposure~7 (Figure~\ref{fig:fors2g07}). Excluding the exposure from the photometry and stellar templates left three of five brackets ($I$ frames~2, 4, and 10), giving $(g-i)_{\rm P1}=+0.691^{+0.173}_{-0.175}$~mag after the artificial-source correction (Appendix~\ref{app:fors2extract}). The November~14 FORS2 and November~11 Keck measurements agree in $(g-r)_{\rm P1}$. Keck gives $0.510^{+0.114}_{-0.119}$~mag after conversion (Appendix~\ref{app:cousins_conversion}). Independent draws give a FORS2-minus-Keck difference of $+0.012$~mag ($-0.113$ to $+0.142$~mag).

The FORS2 $(i-z)_{\rm P1}$ intervals from paired flux draws overlap on November~7 and 15 (Table~\ref{tab:nightphot}). The individual November~7 magnitude medians are $i_{\rm P1}=23.79^{+0.22}_{-0.13}$ and $z_{\rm P1}=23.90^{+0.12}_{-0.09}$~mag.

On November~14 GMOS-N measured $i'-Z=-0.127^{+0.099}_{-0.122}$~mag at the stellar color pivot $(i-z)_{\rm P1}=+0.10$~mag. Propagating the fitted stellar color terms gives $(i-z)_{\rm P1}=-0.224^{+0.145}_{-0.187}$~mag. On November~19 GMOS-N gave PS1-equivalent $r-i=+0.303^{+0.127}_{-0.099}$ and $i-z=-0.105^{+0.132}_{-0.140}$~mag. The $i-z$ color interval includes zero and overlaps the November~14 Gemini color interval. The shared $i$ produces a negative correlation between $r-i$ and $i-z$ (Pearson $\rho=-0.887$) and cancels in their sum. The paired draws give $r-z=+0.198$~mag ($+0.107$ to $+0.285$~mag), a narrower interval than independent color errors would imply.

On November~21 FORS2 recovered \qq\ at the predicted position in both halves of each filter sequence. The full $R_{\rm SPECIAL}$ and $I_{\rm BESS}$ stacks have signal-to-noise ratios of 36.4 and 15.6. The nine brackets yield $(r-i)_{\rm P1}=+0.061^{+0.088}_{-0.083}$~mag, or $(r-i)_{\rm SDSS}=+0.105^{+0.112}_{-0.105}$~mag after converting both apparent magnitudes within each draw. After converting Keck $R_{\rm C}$ to SDSS $r$, independent draws give a FORS2-minus-Keck difference in $r-i$ of $-0.231$~mag ($-0.505$ to $+0.086$~mag).

\subsection{Combined SDSS colors}\label{sec:combinedcolors}

We fitted ten colors from eight visits in their calibrated passbands, retaining the within-visit correlations for Gemini and Keck (Appendix~\ref{app:jointsdss}). The combined SDSS AB colors are $g-r=0.594^{+0.056}_{-0.053}$, $g-i=0.840^{+0.066}_{-0.065}$, and $i-z=-0.104^{+0.067}_{-0.057}$~mag. Adding the November~21 FORS2 data shifts $r-i$ from $0.286^{+0.063}_{-0.058}$ to $0.247^{+0.047}_{-0.057}$~mag and the median $i-z$ by $+0.004$~mag (Table~\ref{tab:nightphot}).

\begin{figure}[!tbp]
\centering
\includegraphics[width=\columnwidth]{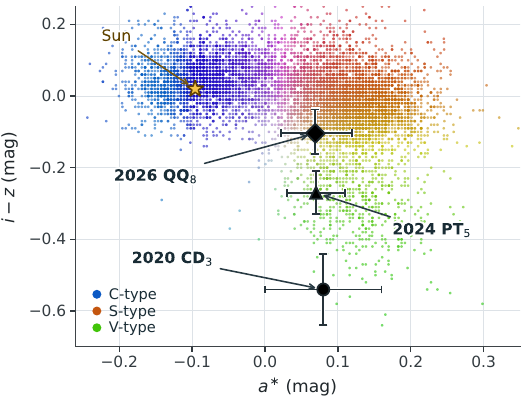}
\caption{Apparent colors of \qq\ (black diamond), \cdthird\ (black circle), \ptfive\ (black triangle), and the CALSPEC Sun (gold star). Bars for \qq\ span the 16th--84th percentiles of the joint November~7--21 SDSS-equivalent AB colors. For \cdthird\ and \ptfive, bars give published $1\sigma$ errors, with LRIS $i-\mathrm{RG850}$ and GMOS $i-Z$ as proxies for $i-z$ \citep{Bolin2020CD3,Bolin2025PT5}. Small points are SDSS Moving Object Catalog asteroids \citep{Ivezic2001}, colored by position in this plane \citep{Ivezic2002,Parker2008}. C types tend to occupy the blue region at negative $a^\ast$, S types the red and orange region at positive $a^\ast$, and V types the green region at positive $a^\ast$ and low $i-z$.}\label{fig:asteroidcolors}
\end{figure}

\subsection{Comparison with asteroid colors}

For \qq, the apparent colors give $a^\ast=0.89(g-r)+0.45(r-i)-0.57=0.069^{+0.050}_{-0.047}$~mag across draws \citep{Ivezic2001}, retaining the incident solar color (Figure~\ref{fig:asteroidcolors}). For \cdthird, \citet{Bolin2020CD3} measured $a^\ast=+0.08\pm0.08$ and $i-\mathrm{RG850}=-0.54\pm0.10$~mag. For \ptfive, $a^\ast=+0.07\pm0.04$ and $i-Z=-0.27\pm0.06$~mag \citep{Bolin2025PT5}.

Subtracting these published red colors from the SDSS $i-z$ of \qq\ gives $+0.440$~mag ($+0.324$ to $+0.557$) for \cdthird\ and $+0.169$~mag ($+0.085$ to $+0.256$) for \ptfive. With independent Gaussian errors on the published colors, these differences are positive in 99.988\% and 97.86\% of draws ($3.67\sigma$ and $2.03\sigma$, one-sided). The red filters differ in their coverage of the $1\,\mu$m absorption. Modeling the four \cdthird\ LRIS magnitudes with the 2020 passbands and its calibration star's catalog colors gives a \qq-minus-\cdthird\ difference of $+0.024$~mag ($-0.075$ to $+0.125$~mag) on common SDSS $i-z$ passbands.

\begin{figure}[!tbp]
\centering
\includegraphics[width=\columnwidth]{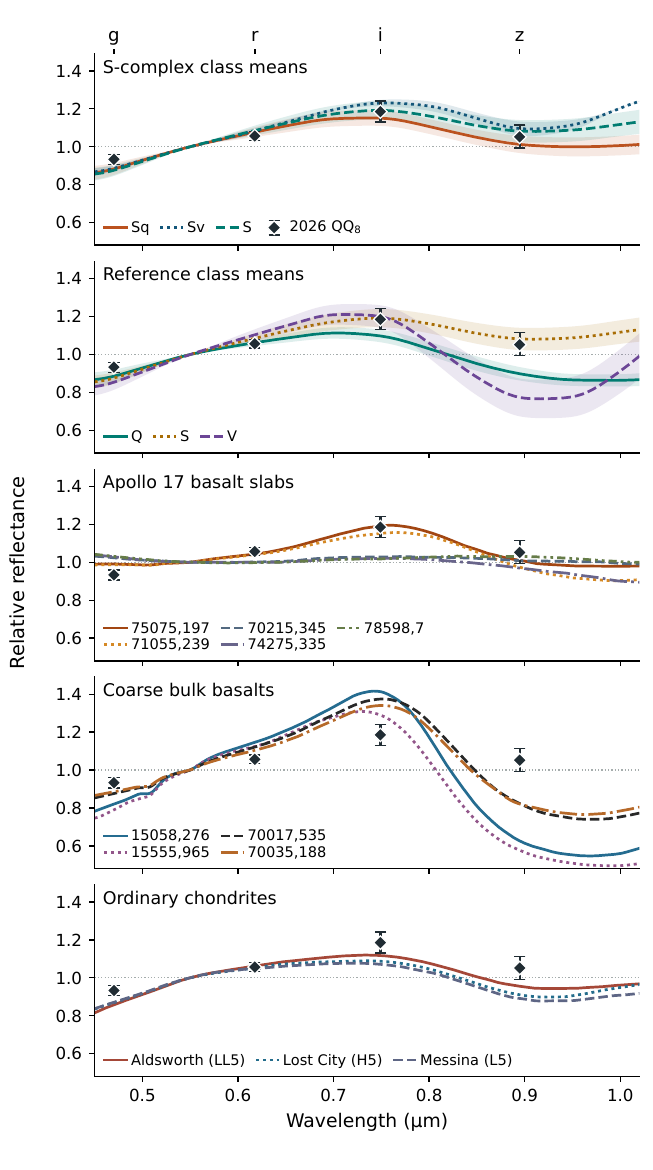}
\caption{Solar-relative reflectance of \qq\ from the combined November~7--21 Keck, Gemini, and VLT photometry. Diamonds and bars give the median and 16th--84th percentiles at the SDSS $g,r,i,z$ pivots (0.470--$0.895\,\mu$m). We normalize each draw to unity at $0.55\,\mu$m. From top to bottom, the curves show the Bus--DeMeo Sq, Sv, and S class means, the Q, S, and V reference means \citep{DeMeo2009}, five intact Apollo~17 basalt slabs \citep{Robertson2022Slabs}, four coarse bulk lunar basalts \citep{Isaacson2011}, and the closest measured meteorite spectrum in each of the LL5, H5, and L5 groups \citep{Dunn2010OC}. Shading shows the dispersion among class members. We fit the unsmoothed spectra to the measured colors (Appendix~\ref{app:reflectspec}). Slab 75075,197 gives the lowest $\chi^2$ at the median colors.}\label{fig:surface}
\end{figure}

For a G2V source, weighting the photon responses by the CALSPEC solar photon spectrum \citep{Colina1996} gives mean detected wavelengths of $0.922\,\mu$m for LRIS RG850, $0.914\,\mu$m for FORS2 Gunn $z$, and $0.880\,\mu$m for GMOS-N $Z$ \citep{Bolin2020CD3,ESOFORS2Filters,Bolin2025PT5,GeminiGMOSFilters}. For RG850 we held transmission constant beyond the last measurement at $0.976\,\mu$m, with the detector response setting the red cutoff. GMOS-N $Z$ spans $0.830$--$0.925\,\mu$m for both \ptfive\ and \qq\ \citep{GeminiGMOSFilters}. We plot the SDSS $z$ point at its $0.895\,\mu$m pivot (Figure~\ref{fig:surface}).

\subsection{Optical reflectance and spectral templates}\label{sec:spectra}

Normalized to unity at $0.55\,\mu$m, the median solar-relative reflectances are 0.933, 1.056, 1.185, and 1.052 in SDSS $g,r,i,z$ (Figure~\ref{fig:surface}). The paired draws give $R_z/R_i=0.884^{+0.056}_{-0.045}$. This $12\%$ decline from $0.749$ to $0.895\,\mu$m is consistent with the blue wing of a silicate absorption band. Its center remains unconstrained.

Of 24 Bus--DeMeo class means \citep{DeMeo2009} and 21 lunar laboratory spectra, Apollo~17 basalt slab 75075,197 fits best ($\chi^2_\nu=0.57$). The Sq, K, and Sv class means all closely match the colors ($\chi^2_\nu=0.87$, 0.90, and 0.97, respectively) (Table~\ref{tab:spectrafits}). For basalt 71055,239, the cut face gives $\chi^2=2.67$, compared with 5.36 for its $>125\,\mu$m particulate fraction \citep{Robertson2022Slabs}.

\begin{table}[!tbp]
\centering\small
\caption{Spectral fits to the combined November 7--21 photometry}\label{tab:spectrafits}
\setlength{\tabcolsep}{3pt}
\begin{tabular}{lrr}
\hline\hline
Spectrum & $\chi^2$ & $\chi^2_\nu$ \\
\hline
\multicolumn{3}{l}{\textit{Apollo 17 cut faces}} \\
75075,197 & 1.71 & 0.57 \\
71055,239 & 2.67 & 0.89 \\
\multicolumn{3}{l}{\textit{Bus--DeMeo class means}} \\
Sq & 2.61 & 0.87 \\
K & 2.70 & 0.90 \\
Sv & 2.92 & 0.97 \\
S & 3.08 & 1.03 \\
Q & 7.48 & 2.49 \\
V & 28.12 & 9.37 \\
\multicolumn{3}{l}{\textit{Ordinary chondrites}} \\
Aldsworth (LL5) & 4.53 & 1.51 \\
Lost City (H5) & 5.57 & 1.86 \\
Messina (L5) & 6.88 & 2.29 \\
\hline\end{tabular}
\par\smallskip\parbox{0.96\columnwidth}{\footnotesize\textit{Note.} Each fit uses the median and covariance of the combined SDSS color draws, with three degrees of freedom. Meteorite rows give the lowest $\chi^2$ in each laboratory group (LL5, H5, L5).}
\end{table}

Across 350,000 joint color draws, the median $\chi^2_{\rm Sq}-\chi^2_{75075,197}$ is $+0.86$ (16th--84th percentiles $-4.08$ to $+5.68$). Slab 75075,197 fits better than Sq in 56.9\% of draws, and Sq fits better than Sv or K in 51.4\% and 49.4\%, respectively. All four spectra remain plausible. For the Earth quasi-satellite (469219) Kamo{\textquoteleft}oalewa, the red slope measured from the ground resembled space-weathered lunar silicates \citep{Sharkey2021Kamo}. Subsequent James Webb Space Telescope spectra showed a shallower slope from 1.0 to $2.5\,\mu$m, consistent with silicate-rich asteroids and leaving its origin uncertain \citep{Sharkey2026Kamo}.

Among the 1,040 near-Earth objects in \citet[][Figures~5 and 12, Section~6.3]{Binzel2019MITHNEOS}, 157 are Sq (15.1\%), ten are K (0.96\%), and three are Sv (0.29\%). Sq is thus about 16 times as common as K and 52 times as common as Sv in this sample, whose median diameter is 0.7~km. If these relative frequencies extend to meter-sized objects, we favor Sq among the asteroid interpretations for \qq.

Of 18 H5, L5, and LL5 ordinary chondrites \citep{Dunn2010OC,Milliken2020RELAB}, Aldsworth (LL5) fits best. Its $R_z/R_i=0.885$ matches the observed decline, although its synthetic $g-r=0.684$~mag is redder than the measured $0.594$~mag. Slab 75075,197 fits better than Aldsworth in 69.3\% of draws. The meteorites were measured as powders, so grain size and texture limit the comparison with exposed rock.

D.~Britt (private communication, 2026 November~18) interpreted the earlier colors as consistent with a fresh ordinary-chondrite-like surface. Space weathering can redden ordinary chondrites and weaken their absorption bands, producing S-like spectra \citep{Binzel2010Weathering,Hasegawa2019Q}. Lunar basalts can likewise redden and lose band contrast \citep{Pieters1993Fines}. Limited retention of fine regolith on a meter-scale body could mute the reddening \citep{Hsu2022Regolith}. For two larger fast rotators, Yarkovsky drift measurements imply low surface thermal conductivity for (499998)~2011~PT and low thermal inertia for 2016~GE$_1$. Insulating regolith could explain both results \citep{Fenucci2021PT,Fenucci2023GE1}. Kamo{\textquoteleft}oalewa's measured drift likewise permits a regolith-covered surface \citep{Fenucci2025KamoThermal}. Typical basalts are expected to show a steeper decline from $0.75$ to $0.90\,\mu$m because the first band lies near the continuum and the second samples the blue wing of the pyroxene absorption near $1\,\mu$m \citep[D.~Britt, private communication;][]{Isaacson2011,Robertson2022Slabs}. Four coarse bulk basalts decline by 31--48\% \citep{Isaacson2011}. The cut faces of 75075,197 and 71055,239 decline by 11.8\% and 12.5\%, close to the measured 12\% (Figures~\ref{fig:surface} and \ref{fig:lunarpreparations}). Cut rock preserves mineral contacts that grinding can disrupt \citep{Robertson2022Slabs}. The colors constrain neither exposure age nor regolith coverage.

\section{Astrometry and orbit determination}\label{sec:dynamics}

The November~22 fit uses 71 positions spanning 93.419~days (August~19--November~21 UTC). Of these, 34 came from other observatories. We matched our 26 VLT, eight Gemini North, and three Keck~I positions to the MPC submissions by identifier to avoid duplicates. We excluded the first two K88 positions on August~19 (22:07:52.723 and 22:11:03.840 UTC) for their large residuals and retained the third (22:14:14.352 UTC) and all later positions. Table~\ref{tab:allastrometry} lists the positions, weights, and observed-minus-computed residuals.

The VLT and Keck submissions appear in MPECs \citetalias{MPEC2026V185}, \citetalias{MPEC2026V273}, and \citetalias{MPEC2026V323}, the November~14 VLT and Gemini positions in \citetalias{MPEC2026V337}, and the November~15 VLT positions in \citetalias{MPEC2026W10}. MPEC~\citetalias{MPEC2026W64} includes the four November~19 Gemini $r'$ positions. Relative to our November~19 fit, we removed 15 unsubmitted Gemini $i'$ and $Z$ positions and added ten submitted November~21 FORS2 $R_{\rm SPECIAL}$ positions. We excluded the four November~21 FORS2 $I$ centroids and the expanded 28-exposure Keck reduction from the orbit fit.

We retained the November~21 positions and midpoints and assigned each Keck exposure an independent 3.0~s timing uncertainty. Our positions use measured east--north error ellipses, timing uncertainties, and shared calibration errors. External positions use the station, catalog, epoch, and nightly multiplicity weights of \citet{Veres2017} (Appendix~\ref{app:orbitweights}).

At Julian Date (JD) 2461303.5 in Barycentric Dynamical Time (TDB, 2026 September~20.0), the heliocentric orbit is nearly circular ($e=0.023$) and inclined by just $0.298^\circ$ to the J2000 ecliptic. The bound geocentric orbit is more eccentric ($e_{\rm geo}=0.595$) and inclined by $37.10^\circ$ to the same plane (Table~\ref{tab:orbitelements}).

\begin{table}[!tbp]\centering
\footnotesize
\setlength{\tabcolsep}{2pt}
\renewcommand{\arraystretch}{0.90}
\caption{Adopted orbital elements of \qq\ with radial SRP\label{tab:orbitelements}}
\begin{tabular}{@{}lr@{\hspace{2pt}}l@{}}\hline\hline Quantity & \multicolumn{2}{c}{Value} \\
\hline
\multicolumn{3}{l}{\textit{Heliocentric elements}} \\
\noalign{\vskip 0.6pt}
\hline
\noalign{\vskip 0.6pt}
Epoch (JD TDB) & 2461303.5 &  \\
\noalign{\vskip 0.6pt}
\hline
\noalign{\vskip 0.6pt}
Perihelion time, $T_p$ (JD TDB) & $2460987.637497$ & $\pm\,(6.4\times10^{-5})$ \\
Semimajor axis, $a$ (au) & $1.015364283$ & $\pm\,(3.3\times10^{-8})$ \\
Eccentricity, $e$ & $0.022898278$ & $\pm\,(2.1\times10^{-8})$ \\
Perihelion, $q$ (au) & $0.992114190$ & $\pm\,(1.5\times10^{-8})$ \\
Aphelion, $Q$ (au) & $1.038614376$ & $\pm\,(5.4\times10^{-8})$ \\
Inclination, $i$ (deg) & $0.29789525$ & $\pm\,(3.4\times10^{-7})$ \\
Ascending node, $\Omega$ (deg) & $211.850427$ & $\pm\,(5.9\times10^{-5})$ \\
Arg. of perihelion, $\omega$ (deg) & $203.332598$ & $\pm\,(7.8\times10^{-5})$ \\
Mean anomaly, $M$ (deg) & $304.277138$ & $\pm\,(4.7\times10^{-5})$ \\
\noalign{\vskip 0.6pt}
\hline
\noalign{\vskip 0.6pt}
\multicolumn{3}{l}{\textit{Geocentric elements}} \\
\noalign{\vskip 0.6pt}
\hline
\noalign{\vskip 0.6pt}
Epoch (JD TDB) & 2461303.5 &  \\
\noalign{\vskip 0.6pt}
\hline
\noalign{\vskip 0.6pt}
Perigee time, $T_{p,{\rm geo}}$ (JD TDB) & $2461267.8141505$ & $\pm\,(9.3\times10^{-6})$ \\
Semimajor axis, $a_{\rm geo}$ (au) & $0.009742854$ & $\pm\,(1.1\times10^{-8})$ \\
Eccentricity, $e_{\rm geo}$ & $0.59466557$ & $\pm\,(5.6\times10^{-7})$ \\
Perigee, $q_{\rm geo}$ (au) & $0.0039491143$ & $\pm\,(9.2\times10^{-9})$ \\
Apogee, $Q_{\rm geo}$ (au) & $0.015536594$ & $\pm\,(1.3\times10^{-8})$ \\
Inclination, $i_{\rm geo}$ (deg) & $37.096007$ & $\pm\,(4.4\times10^{-5})$ \\
Ascending node, $\Omega_{\rm geo}$ (deg) & $300.178870$ & $\pm\,(3.6\times10^{-5})$ \\
Arg. of perigee, $\omega_{\rm geo}$ (deg) & $19.97988$ & $\pm\,(1.0\times10^{-4})$ \\
Mean anomaly, $M_{\rm geo}$ (deg) & $63.38462$ & $\pm\,(1.0\times10^{-4})$ \\
\noalign{\vskip 0.6pt}
\hline
\noalign{\vskip 0.6pt}
\multicolumn{3}{l}{\textit{Radial force and fit statistics}} \\
\noalign{\vskip 0.6pt}
\hline
\noalign{\vskip 0.6pt}
Radial SRP, $A_1$ (au day$^{-2}$) & $1.363\times10^{-10}$ & $\pm\,(9.0\times10^{-12})$ \\
Effective $C_{\rm R} A/m$ (m$^2$ kg$^{-1}$) & $6.016\times10^{-4}$ & $\pm\,(3.97\times10^{-5})$ \\
$\chi^2$ / d.o.f. & 43.50 / 135 &  \\
Absolute magnitude, $H$ (mag) & $31.37$ & $\pm\,0.34$
\\ \hline\end{tabular}
\par\smallskip\parbox{0.98\columnwidth}{\footnotesize\textit{Note.} Elements refer to the J2000 ecliptic at 2026 September~20, 00:00 TDB. The periapsis times give the most recent passage on each osculating ellipse. The formal $1\sigma$ errors propagate the Cartesian fit covariance with the DE440s Earth state held fixed. For solar irradiance of 1361~W~m$^{-2}$ at 1~au, the effective ratio equals the geometric $A/m$ for $C_{\rm R}=1$. JPL Horizons solution~7 gives $H=31.367$~mag for $G=0.150$ \citep{JPLHorizonsQQ8}. The corresponding Small-Body Database record gives $\sigma_H=0.34247$~mag \citep{JPLSBDBQQ8}.}
\end{table}

\subsection{Dynamical source regions}\label{sec:neosources}

The near-Earth object population model (NEOMOD) combines orbital integrations with Catalina Sky Survey observations to estimate source fractions by orbit and absolute magnitude \citep{Nesvorny2023NEOMOD}. NEOMOD3 adopts the NEOMOD2 source distributions and omits lunar ejecta \citep{Nesvorny2024NEOMOD2,Nesvorny2024NEOMOD3}. For Kamo{\textquoteleft}oalewa, population models favor main-belt delivery over ejection from the lunar crater Giordano Bruno \citep{Fenucci2026KamoOrigin}. At $H=31.37$~mag (Table~\ref{tab:orbitelements}), \qq\ is fainter than the last bin ($H=27.75$--$28.00$~mag). We therefore report fractions for modeled analogues at its center ($H=27.875$~mag).

We propagated the nominal SRP orbit back 365~days with IAS15 and the DE440 and SB441-N16 ephemerides. At JD~2460938.5 TDB the pre-encounter elements were $(a,e,i)=(0.94276~\mathrm{au},0.05419,0.3173^\circ)$. In the published NEOMOD3 grid \citep{Nesvorny2024NEOMOD3}, this cell assigns 78.8\% to $\nu_6$, 20.6\% to Jupiter's 3:1 resonance, 0.2\% to the 5:2 resonance, 0.3\% to weak inner-belt resonances, and 0.1\% to the Hungarias. During capture (JD~2461303.5 TDB) the same magnitude bin gives 89.1\% for $\nu_6$, 10.6\% for 3:1, 0.2\% for weak inner-belt resonances, and 0.1\% for the Hungarias. The November~21 timing solution selected the same two cells.

The young Datura family contains subkilometer inner-belt fragments \citep{Vokrouhlicky2017Datura}, and other young family breakups have been linked to ordinary-chondrite falls \citep{Broz2024Meteorites}. Diurnal Yarkovsky drift moves prograde fragments outward and retrograde fragments inward, as observed in the Eos family \citep{Hanus2018Eos}. In the inner belt, outward drift can feed Jupiter's 3:1 resonance from its sunward side and inward drift can feed $\nu_6$ from larger semimajor axes. Either resonance can raise eccentricity enough to reach near-Earth space \citep{Nugent2012Drift,Nesvorny2023NEOMOD}. The parent family and spin state of \qq\ are unknown.

\section{Solar radiation pressure and bulk density}\label{sec:physical}

\subsection{The measured radial acceleration}\label{sec:srp}

The outward acceleration at 1~au is $A_1=(1.36285\pm0.08985)\times10^{-10}$~au~day$^{-2}$ (formal significance $15.17\sigma$). Including $A_1$ reduces the unweighted root-mean-square (rms) sky-plane residual from $1\farcs345$ to $0\farcs374$ ($0\farcs951$ to $0\farcs264$ per coordinate). Tables~\ref{tab:nov21-fit-summary} and \ref{tab:gravity-refit} give the fit statistics and independent six-parameter gravity solution.

With the revised weights, the 61 earlier positions give $A_1=(1.370\pm0.097)\times10^{-10}$~au~day$^{-2}$ and predicts the ten November~21 $R$ positions with a radial rms of $0\farcs116$, close to their postfit rms of $0\farcs112$. Under gravity alone, the corresponding residuals are $0\farcs401$ and $0\farcs259$. The earlier JPL solution~7 gave $A_1=(1.306\pm0.100)\times10^{-10}$~au~day$^{-2}$ from 63 public positions through November~19 \citep{JPLSBDBQQ8}.

At a solar irradiance of 1361~W~m$^{-2}$ at 1~au \citep{Prsa2016}, the measured $A_1$ implies $\mathcal A=C_{\rm R}A/m=(6.016\pm0.397)\times10^{-4}$~m$^2$~kg$^{-1}$ for radiation pressure acting on projected area $A$ and mass $m$. We adopt a radiation-pressure coefficient $C_{\rm R}=1$. The inferred $A/m$ and its formal uncertainty scale as $C_{\rm R}^{-1}$.

\subsection{Diameter, bulk density, and mass}\label{sec:density}

We adopt $H=31.37\pm0.34$~mag from JPL Horizons ($G=0.15$) \citep{JPLHorizonsQQ8}, with the uncertainty from the JPL Small-Body Database \citep{JPLSBDBQQ8}. The diameter--magnitude relation gives $D\sqrt{p_V}=(1329\,{\rm km})\,10^{-H/5}=0.708^{+0.120}_{-0.103}$~m \citep{Pravec2012WISE}. Because the reflectances are normalized, we estimate the visible geometric albedo $p_V$ from spectral and dynamical analogues. For a sphere of diameter $D$ and effective area-to-mass ratio $\mathcal{A}=C_{\rm R}A/m$, the bulk density and mass satisfy $\rho/C_{\rm R}=3/(2D\mathcal{A})$ and $m/C_{\rm R}=\pi D^2/(4\mathcal{A})$. At fixed $D$ and $\mathcal A$, density and mass scale linearly with $C_{\rm R}$. For $C_{\rm R}=1$ we sample $H$ and $A_1$ from independent Gaussians, truncating $A_1$ at zero. We pair the same draws with each independently chosen albedo and report medians and 16th--84th percentiles.

Six optically selected Sq asteroids have median $p_V=0.243$ and object-to-object standard deviation 0.039 \citep{Mainzer2011Sq}. If they represent meter-scale Sq surfaces, a Gaussian albedo distribution with those parameters, truncated at zero, gives $D=1.44^{+0.28}_{-0.23}$~m, $\rho=1.73^{+0.36}_{-0.30}$~g~cm$^{-3}$, and $m=(2.73^{+1.19}_{-0.82})\times10^3$~kg.

Adopting the NEOMOD3 mean $p_V=0.191$ for 10--100~m NEOs \citep[][Figure~14]{Nesvorny2024NEOMOD3} gives $D=1.62^{+0.28}_{-0.24}$~m, $\rho=1.54^{+0.29}_{-0.24}$~g~cm$^{-3}$, and $m=(3.43^{+1.30}_{-0.94})\times10^3$~kg. The model extrapolates its thermal calibration below about 100~m, and \qq\ falls below its 10--100~m bin.

Apollo~17 slabs 75075,197 and 71055,239 have solar-weighted Johnson $V$ reflectance factors of 0.096 and 0.148 at $30^\circ$ phase \citep{Robertson2022Slabs}. For a Lambert sphere these imply geometric albedos of 0.064 and 0.098 (Appendix~\ref{app:slabalbedo}). Their mean, $p_V=0.081$, gives $D=2.49^{+0.42}_{-0.36}$~m, $\rho=1.00^{+0.19}_{-0.16}$~g~cm$^{-3}$, and $m=(8.09^{+3.06}_{-2.22})\times10^3$~kg. Separately the slabs give $D=(2.80,2.26)$~m, $\rho=(0.89,1.11)$~g~cm$^{-3}$, and $m=(10.28,6.67)\times10^3$~kg. A Lommel--Seeliger sphere with a constant surface phase function gives $D=(2.37,1.91)$~m, $\rho=(1.05,1.31)$~g~cm$^{-3}$, and $m=(7.35,4.76)\times10^3$~kg in the same slab order. The intervals exclude uncertainty from opposition brightening, which a single laboratory phase angle cannot constrain.

Assuming main-belt delivery, we weighted the albedo distributions of \citet[][Table~1]{Morbidelli2020Albedo} by the nominal pre-encounter source fractions (Section~\ref{sec:neosources}). We grouped the weak inner-belt resonances with $\nu_6$, 8:3 and 7:3 with 5:2, and 9:4 and 11:5 with 2:1. The Hungarias, Phocaeas, 3:1 resonance, and Jupiter-family comets retained their assignments. The grouped weights (79.1\% $\nu_6$, 20.6\% 3:1, 0.2\% 5:2, and 0.1\% Hungaria) give probabilities of 12.5\% for $p_V\leq0.10$, 60.4\% for $0.10<p_V\leq0.30$, and 27.1\% for $p_V>0.30$.

The albedo probability density is uniform within the first two bins and declines exponentially in the third \citep[Equation~4]{Morbidelli2020Albedo}. The median $p_V$ is 0.224 (mean 0.236). Its 16th--84th percentile range of 0.112--0.355 includes the Sq median and NEOMOD3 mean. Propagating this distribution with $H$ and $A_1$ gives $D=1.52^{+0.71}_{-0.39}$~m, $\rho=1.64^{+0.58}_{-0.53}$~g~cm$^{-3}$, and $m=(3.03^{+3.54}_{-1.36})\times10^3$~kg at fixed source weights and albedo-class probabilities. This estimate extrapolates beyond the source grid and the $16<H<24$ range in which \citet{Morbidelli2020Albedo} tested the dark fraction for $\nu_6$.

\begin{figure*}[!t]
\centering
\includegraphics[width=.92\textwidth]{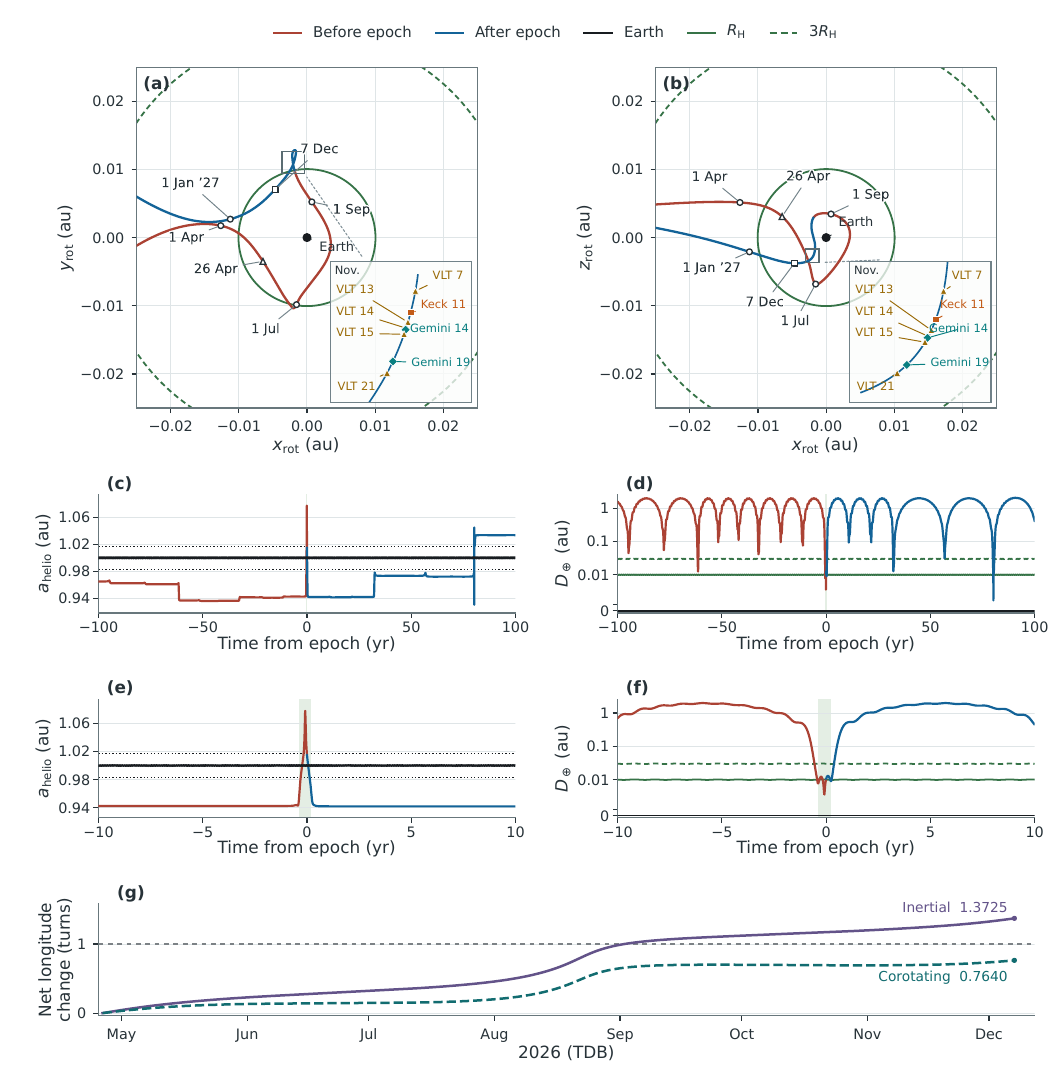}
\caption{Earth encounter of \qq\ on the nominal November~22 fit to 71 positions with radial solar radiation pressure. (a,b) Earth-centered projections over $\pm1$~yr about 2026 September~20.0 TDB. In the Sun--Earth corotating frame, positive $x$ points away from the Sun and positive $z$ north of the J2000 ecliptic. Red precedes the epoch and blue follows it. The open triangle marks the nominal capture entry on April~26 at 07:40:31 TDB and the open square marks the predicted exit on December~7 at 06:40:54 TDB. Insets mark the mean exposure midpoints of the eight VLT (triangle), Keck (square), and Gemini (diamond) visits, using the 19 clean frames on November~19. Solid and dashed green curves mark one and three Earth Hill radii, evaluated at the epoch in (a,b) and at each time in (d,f). Heliocentric semimajor axis and geocentric distance span $\pm100$~yr in (c,d) and $\pm10$~yr in (e,f). Black curves mark Earth's semimajor axis and zero geocentric distance, dotted lines Earth's epoch perihelion and aphelion, and green shading the capture interval. The distance scale is linear below $0.003$~au and logarithmic above. (g) Net geocentric longitude change since capture entry in the inertial and Sun--Earth corotating ecliptic frames.}
\label{fig:encounter}
\end{figure*}

\begin{figure*}[!t]
\centering
\includegraphics[width=.94\textwidth]{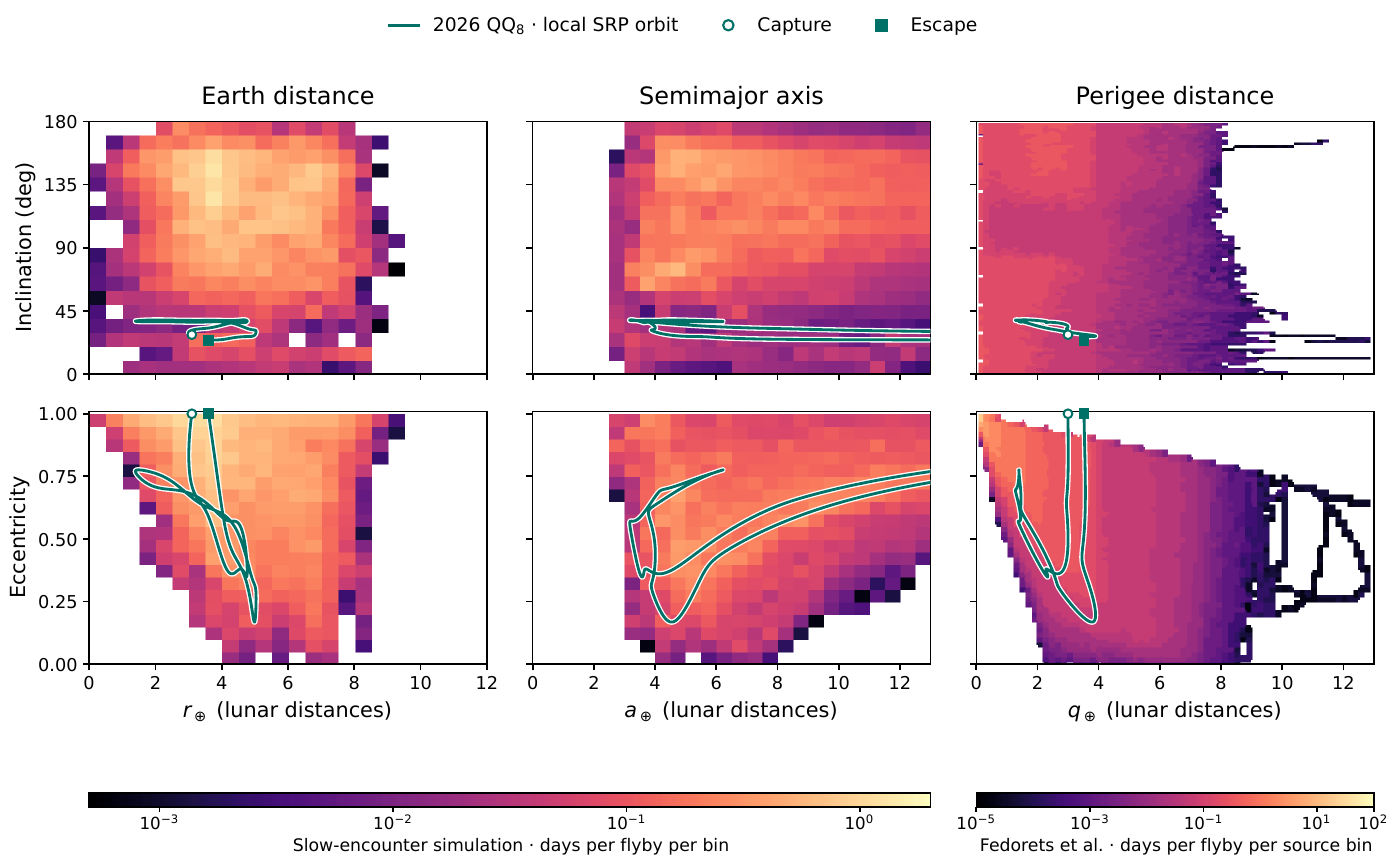}
\caption{Geocentric inclination to the J2000 ecliptic (top) and eccentricity (bottom) during the 2026 capture of \qq. The curve samples the November~22 nominal SRP orbit every 3~h, with both energy crossings evaluated at their refined times. The horizontal axes give Earth distance $r_\oplus$, osculating semimajor axis $a_\oplus$, and osculating perigee $q_\oplus$, in lunar distances (384,400~km). In the distance and perigee panels, open circles mark capture on April~26 and filled squares mark the predicted escape on December~7. At both energy crossings $e_\oplus=1$ and $a_\oplus$ diverges. The left and middle columns show mean days per bin for 378 completed simulated slow-encounter flybys on a common scale. Bins span 0.5 lunar distances and either $10^\circ$ in inclination or 0.05 in eccentricity. The right column gives approximate mean days per original bin on a separate scale, digitized from \citet[][Figure~10]{Fedorets2017} with the perigee-axis correction of \citet{Naidu2021}.}
\label{fig:flybys}
\end{figure*}

\section{Capture and long-term orbital evolution}

\subsection{The 2026 capture}\label{sec:capture}

For all integration windows we used the same 1,000 draws from the seven-parameter fit covariance. Each pair shared its initial heliocentric position and velocity. One member retained the sampled $A_1$ at 1~au, scaled with inverse-square heliocentric distance. Its gravity-only sibling had $A_1=0$. The independent gravity fit serves only for the astrometric comparison (Section~\ref{sec:srp}).

We define capture by negative geocentric two-body energy within three Earth Hill radii, with the Hill radius evaluated at Earth's heliocentric distance (Equation~\ref{eq:qq8-capture-energy-hill}) \citep{Granvik2012,Fedorets2017}. All clones passed within one Hill radius while bound. With SRP, the median closest approach on August~18 was 1.3971 lunar distances ($0.355R_{\rm H}$). Median capture extended from April~26 at 07:40 to a predicted exit on December~7 at 06:41 TDB, a duration of $224.95872^{+0.00107}_{-0.00134}$~days (Table~\ref{tab:captureuncertainty}). Figure~\ref{fig:encounter} locates the observations along the nominal path. Within matched pairs, SRP advanced entry by a median of 23.54~min, delayed escape by 4.11~min, and lengthened capture by 27.66~min (16th--84th percentiles, 25.91--29.43~min). Across 10,000 bootstrap resamples of intact pairs, the respective Monte Carlo standard errors are 0.07, 0.01, and 0.08~min.

In the Sun--Earth corotating frame the median net change in geocentric longitude was 0.76399 turns with SRP and 0.76378 under gravity alone (0.78282 and 0.78261 when summing absolute increments). All clones qualify as flybys by corotating longitude \citep{Granvik2012,Fedorets2017,Jedicke2025}. The tangent to the three-dimensional corotating path turned through 1.32923 and 1.32907 revolutions, respectively, placing both ensembles in the orbiter class under the intrinsic-curvature criterion \citep{Urrutxua2017,Jedicke2018}. Table~\ref{tab:captureuncertainty} also gives the longitude change in inertial ecliptic coordinates.

\begin{figure*}[!t]
\centering
\includegraphics[width=.91\textwidth]{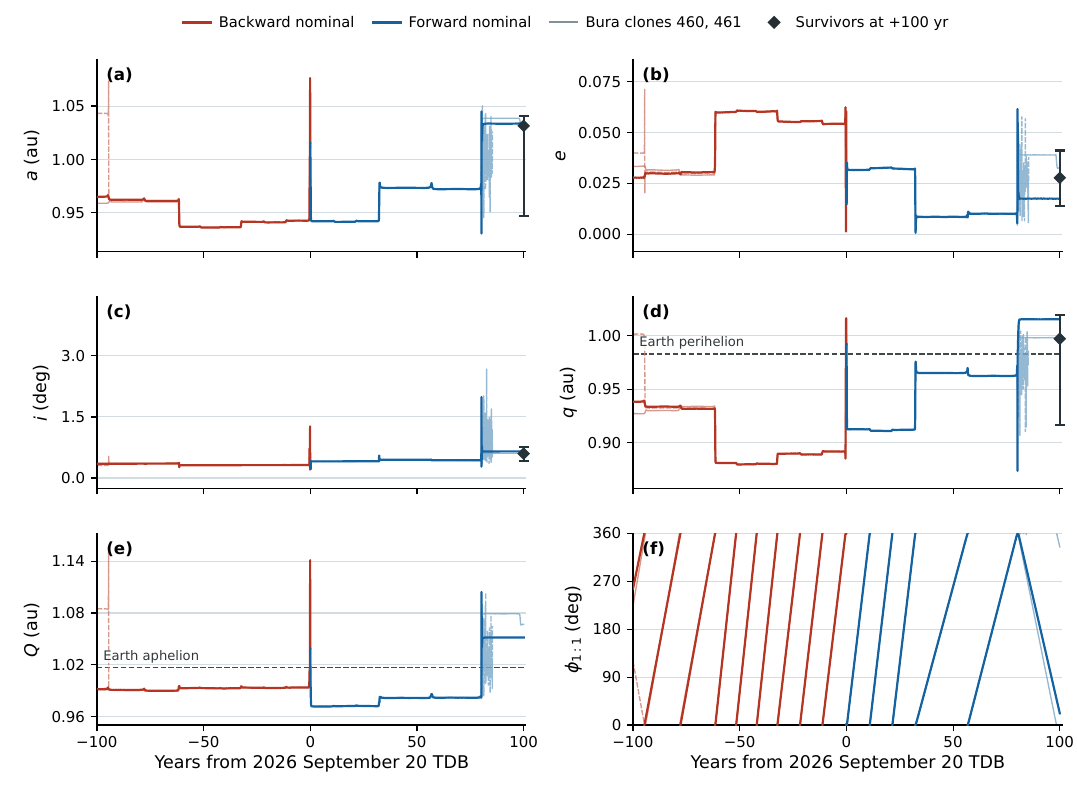}
\caption{Heliocentric evolution of the November~22 orbit fit (71 positions) under gravity and radial solar radiation pressure over $\pm100$~yr from JD~2461303.5 TDB. Panels (a)--(e) give semimajor axis, eccentricity, inclination to the J2000 ecliptic, perihelion, and aphelion. Panel~(f) gives the Earth 1:1 resonant angle $\phi=\lambda_{\rm QQ8}-\lambda_\oplus$, where $\lambda=M+\omega+\Omega$ is heliocentric mean longitude. Curves break at $0^\circ/360^\circ$. Red and blue mark times before and after the epoch. Heavy curves sample a separate integration of the same nominal 71-position orbit every five days. Light curves show the two archived Bura clone histories at 30-day intervals. Clone~460 (dashed) struck Earth at $+85.4545$~yr, and clone~461 (solid) reached $+100$~yr. Dashed horizontal lines in (d,e) give Earth's perihelion and aphelion at the epoch. In (a)--(d), diamonds and bars give the medians and 16th--84th percentiles of all 994 SRP survivors at $+100$~yr.}\label{fig:centuryelements}
\end{figure*}

During nominal SRP capture, \qq\ spans 1.40--5.04 lunar distances from Earth, with osculating geocentric perigee of 1.28--3.90 lunar distances and inclination of $24.1^\circ$--$38.6^\circ$ to the J2000 ecliptic (Figure~\ref{fig:flybys}). Our slow-encounter simulation yielded 378 flybys from 131,072 inward crossings at 0.0325~au and geocentric speeds of 0.45--1.0~km~s$^{-1}$. For comparison at perigee we use the distribution of \citet{Fedorets2017} with the axis correction of \citet{Naidu2021}.

\subsection{Evolution through subsequent Earth encounters}\label{sec:century}

Over $\pm100$~yr the nominal heliocentric 1:1 resonant angle accumulates a net 8.29 turns before the epoch and 4.06 after it (Figure~\ref{fig:centuryelements}f). In both ensembles the preceding bound episode ends a median of 61.5~yr before the epoch and the next begins 32.3~yr after it. All clones remain outside one Hill radius during those two episodes. All 1,000 pairs survive to $\pm5$, $\pm10$, $\pm20$, and $\pm50$~yr. At $-5$, $-10$, and $-20$~yr their median three-dimensional separations are 25,705, 64,312, and 17,328~km, compared with 4,887, 10,941, and 12,889~km at the corresponding forward epochs. At $-50$ and $+50$~yr the medians reach 0.01920 and 0.00718~au, respectively.

\begin{figure*}[!tb]
\centering
\includegraphics[width=.94\textwidth]{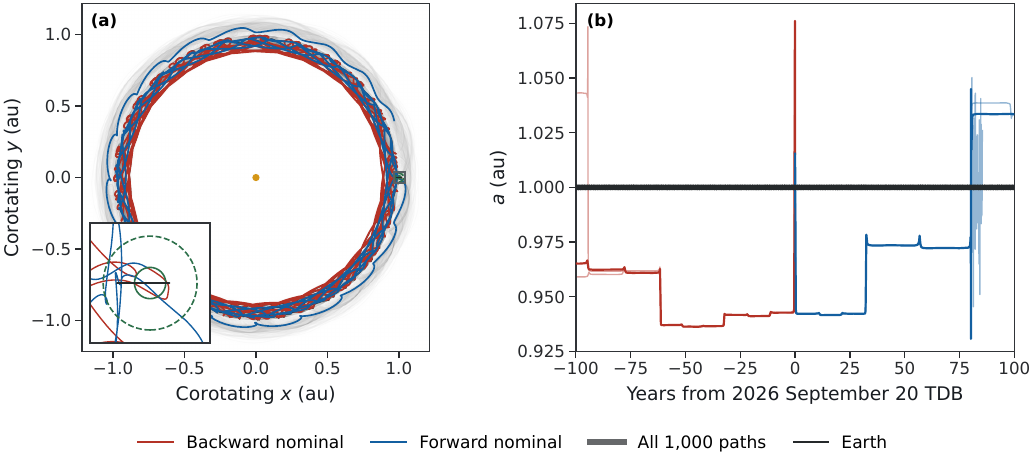}
\caption{Century-scale evolution under gravity and radial solar radiation pressure for the November~22 fit to 71 positions. (a) Heliocentric paths in the plane rotating with Earth's longitude about the J2000 ecliptic pole. Darker pixels contain paths of more of the 1,000 Bura clones, counting each clone once per pixel across both time directions. Sampling is daily and every three hours within one year of the epoch, through first surface contact or the integration limit. Red and blue trace the nominal orbit of Figure~\ref{fig:centuryelements} before and after JD~2461303.5 TDB. Gold marks the Sun. The inset spans 0.077~au and shows the nominal path near Earth, Earth's orbit (black), and one and three Hill radii at the epoch (solid and dashed green). (b) Heliocentric semimajor axes of \qq\ (red and blue) and Earth (black). The nominal and Earth curves are sampled every five days. Pale curves show two Bura clones sampled every 30 days. Clone~460 struck Earth at $+85.4545$~yr, and clone~461 reached $+100$~yr.}\label{fig:centurycorot}
\end{figure*}

At $-100$~yr, 999 gravity controls and 997 SRP clones survive, a paired survival difference (SRP minus gravity) of $-0.20\pm0.20$ percentage points (Monte Carlo standard error). The 996 joint survivors have a median separation of 1.078~au (16th--84th percentiles, 0.322--1.827~au). At $+100$~yr, six SRP clones have struck Earth and all 1,000 gravity controls survive, giving a paired difference of $-0.60\pm0.24$ percentage points. The 994 surviving pairs lie a median of 1.020~au apart (0.467--1.679~au). Figure~\ref{fig:centurycorot} shows the paths of all 1,000 SRP clones in the Earth-corotating plane and the changes in heliocentric semimajor axis during Earth encounters.

At $+100$~yr, the SRP survivors have median $a=1.03150$~au (16th--84th percentiles, 0.94701--1.04045~au), $e=0.02776$ (0.01402--0.04116), $i=0.598^\circ$ ($0.426^\circ$--$0.761^\circ$), and $q=0.99716$~au (0.91683--1.01956~au). The gravity controls give $a=0.97408$~au (0.973925--0.974210~au). SRP broadens the central interval in $a$ by a factor of about 328, from 0.000285 to 0.09344~au.

\begin{figure*}[!t]
\centering
\includegraphics[width=.91\textwidth]{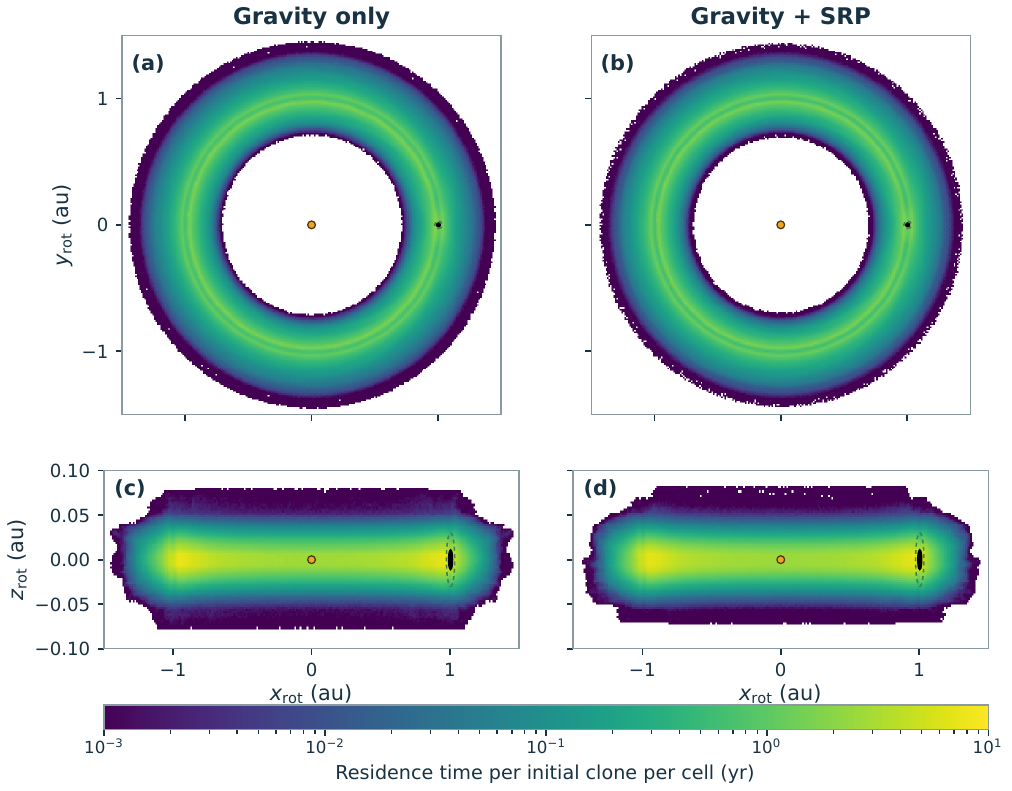}
\caption{Residence of the 1,000 matched pairs from the 71-position November~22 fit over the next 10,000~yr. Left and right columns show gravity alone and gravity with radiation pressure in the Sun-centered corotating frame of Figure~\ref{fig:centurycorot}. Gold marks the Sun. The common logarithmic scale gives years per initial clone per cell. Each clone contributes until Earth or Moon contact or the end of the integration. Cells span $0.0125\times0.0125$~au in $x$--$y$ and $0.0125\times0.0025$~au in $x$--$z$. White denotes zero residence. Daily samples through year~100 and 16-day samples thereafter give 8.508 million clone-years under gravity alone and 8.652 million with radiation pressure in each projection. Solid black and dashed green curves mark one and three Earth Hill radii ($R_{\rm H}=0.01005$~au) at the fit epoch.}\label{fig:tenkyr}
\end{figure*}

\subsection{Forward evolution over 10,000 years}\label{sec:forward10k}

By $+10{,}000$~yr, 791 of 1,000 gravity controls survived, 208 struck Earth, and one struck the Moon. With SRP, 815 survived and 185 struck Earth (Table~\ref{tab:fit-outcomes}). Only the SRP member survived in 173 pairs and only the gravity control in 149. The paired survival difference was $+2.40\pm1.79$ percentage points (Monte Carlo standard error). The approximate 95\% Monte Carlo interval of $-1.12$ to $+5.92$ percentage points spans zero for the adopted orbit, covariance, and force models (Appendix~\ref{app:longintegration}).

In the Earth-corotating ecliptic projection, the central 95\% of the integrated residence time falls at projected heliocentric radii of 0.837--1.213~au under gravity alone and 0.840--1.213~au with SRP (Figure~\ref{fig:tenkyr}). The 95th percentile of $|z|$ lies in the same bin for both models, centered at 0.02625~au. The supplementary atlas shows the projected distance traveled in each cell and individual paths over the final decade.

The 873 pairs surviving to $+1{,}000$~yr lay a median of 1.357~au apart. By $+10{,}000$~yr the median three-dimensional separation of the 642 joint survivors reached 1.348~au (16th--84th percentiles, 0.547--1.916~au, Table~\ref{tab:fit-separation}). Despite this divergence, the central 68\% widths in semimajor axis remained similar, 0.201~au for gravity and 0.198~au for SRP.

For the historical orbit solutions, paired survival differences at the same elapsed horizon were $+1.10\pm1.73$ percentage points for the November~19 fit (V83, 76 positions), $+3.60\pm1.76$ for November~15 (57 positions), and $+0.30\pm1.74$ for JPL solution~6 (59 positions). Tables~\ref{tab:fit-outcomes} and \ref{tab:fit-separation} measure time from each fit's epoch, eight days earlier for November~15, and pair clone IDs only within each fit.

\section{Conclusions}\label{sec:discussion}

The colors, including the 12\% decline in solar-relative reflectance from $i$ to $z$, are consistent with both the Sq asteroid template and exposed Apollo~17 basalt. Slab 75075,197 fits better than Sq in 56.9\% of the color draws. Spectroscopy across the 1 and 2~$\mu$m bands could distinguish exposed basalt from a fresh ordinary-chondrite-like surface.

The $15.2\sigma$ detection of outward acceleration constrains bulk density once an albedo is assumed. For a sphere with $H=31.37\pm0.34$~mag and $C_{\rm R}=1$, the Sq albedo gives a median diameter of 1.44~m and density of 1.73~g~cm$^{-3}$. The darker lunar slabs imply a larger diameter (2.49~m) and lower density (1.00~g~cm$^{-3}$) at their mean albedo (Section~\ref{sec:density}).

During the 224.96-day nominal capture with SRP, \qq\ is a flyby under the corotating-longitude criterion (0.7640 turns) and an orbiter under the three-dimensional path-curvature criterion (1.3292 turns in the same frame). Across 1,000 matched pairs, SRP lengthens capture by a median of 27.66~min (16th--84th percentiles, 25.91--29.43~min). After 10,000~yr, 815 SRP clones and 791 gravity-only siblings survive. The paired difference is $+2.40\pm1.79$ percentage points (Monte Carlo standard error). Its approximate 95\% interval, $-1.12$ to $+5.92$ percentage points, includes zero. The 642 pairs that survive in both models end a median of 1.348~au apart (0.547--1.916~au, 16th--84th percentiles).

\begin{acknowledgments}
\linespread{0.92}\selectfont
Based on observations collected at the European Southern Observatory under ESO programme 117.2C7T.001. Henri M.\,J. Boffin supported preparation of the FORS2 observing blocks. Anni Kasikov and Elyar Sedaghati were the UT1 night astronomers on November~13 UTC, and Sedaghati was also on duty on November~14. Boris Häußler observed \qq\ on November~15.

Leo Y. Alcorn and Bryce T. Bolin co-led the Keck observations under target-of-opportunity program 2026B\_K450 (PI Alcorn). We thank the scheduled Caltech team (C364, PI M. Brown) for accommodating the sequence and Matt Wahl, Matthew Belyakov, and Ian Wong for helping with the Keck I/LRIS observations. The W.\,M. Keck Observatory is a private nonprofit scientific partnership of Caltech, the University of California, and the National Aeronautics and Space Administration (NASA), established with support from the W.\,M. Keck Foundation. After public release, the raw data will be available through the Keck Observatory Archive, operated by Keck and the NASA Exoplanet Science Institute under contract with NASA.

The Gemini North observations were obtained under Director's Discretionary program GN-2026B-DD-108. The International Gemini Observatory is a program of the National Science Foundation (NSF) National Optical-Infrared Astronomy Research Laboratory (NOIRLab). The Association of Universities for Research in Astronomy operates NOIRLab through a cooperative agreement with the U.S. National Science Foundation on behalf of the Gemini partners. The partners are Argentina (Ministerio de Ciencia, Tecnolog\'ia e Innovaci\'on), Brazil (Minist\'erio da Ci\^encia, Tecnologia, Inova\c{c}\~oes e Comunica\c{c}\~oes), Canada (National Research Council), Chile (Agencia Nacional de Investigaci\'on y Desarrollo), the Republic of Korea (Korea Astronomy and Space Science Institute), and the United States (NSF). Brian C. Lemaux and Eunchong Kim served as contact scientists. Mat\'ias Rodr\'iguez-Henr\'iquez obtained the November~14 UTC data with observing support from Jennie L. Berghuis. For November~19, Lemaux and Andrew W. Stephens identified a guide star for the full sequence. Lemaux scheduled the visit, and Leila Alamos obtained the data with System Support Associate Jen Miller. We used calibration data from the Gemini Observatory Archive \citep{Hirst2017GeminiArchive} and reduced the GMOS-N images with Data Reduction for Astronomy from Gemini Observatory North and South (DRAGONS) v4.2.2 \citep{Labrie2023DRAGONS,Simpson2026DRAGONS}.

We recognize the deep cultural and spiritual significance of Maunakea to the K\=anaka Maoli (Native Hawaiian people) and are grateful for the opportunity to observe there with Keck~I and Gemini North. We thank James Westover for arranging access to the Bura cluster and Gordan Jane\v{s} for assistance with the orbital integrations.

We thank T.~Marshall Eubanks, Sam Deen, and Bill J.~Gray for discussing Sar2901's astrometry on the Minor Planet Mailing List before designation. We are grateful to Prof.~Daniel Britt for discussing how the silicate spectra compare with intact lunar rock and to Dr.~Robert Jedicke for clarifying the orbital classification of temporarily captured bodies.

We used generative artificial intelligence (AI) tools (OpenAI's ChatGPT and Codex) to help reduce the imaging data, prepare the figures, and outline the analysis.
\end{acknowledgments}
\facilities{VLT:Antu (FORS2), Keck:I (LRIS), Gemini:Gillett (GMOS-N)}
\software{REBOUND \citep{ReinLiu2012,ReinSpiegel2015}, ASSIST \citep{ASSIST}}
\appendix
\renewcommand{\theHequation}{\thesection.\arabic{equation}}
\section{Observing log}\label{app:obslog}
\startlongtable
\begin{deluxetable}{c l l c r r c r r r r}
\tabletypesize{\footnotesize}
\setlength{\tabcolsep}{3.1pt}
\tablecaption{VLT/FORS2, Keck/LRIS, and Gemini/GMOS-N observations of \qq\ by visit and filter\label{tab:obslog}}
\tablehead{\colhead{Nov.} & \colhead{Instr.} & \colhead{Filter} & \colhead{UTC midpoints} & \colhead{Frames} & \colhead{Integration} & \colhead{Airmass} & \colhead{FWHM} & \colhead{$r_{\rm h}$} & \colhead{$\Delta$} & \colhead{$\alpha$} \\
\colhead{(2026)} & \colhead{} & \colhead{} & \colhead{(hh:mm)} & \colhead{} & \colhead{(s)} & \colhead{} & \colhead{($\arcsec$)} & \colhead{(au)} & \colhead{(au)} & \colhead{(deg)}}
\startdata
7 & FORS2 & $I_{\rm BESS}$ & 07:52--08:46 & 10 & 600 & 1.40--1.75 & 0.65 & 0.9895 & 0.01224 & 97.3 \\
7 & FORS2 & $z_{\rm GUNN}$ & 07:55--08:49 & 10 & 1200 & 1.39--1.73 & 0.66 & 0.9895 & 0.01224 & 97.3 \\
11 & LRIS & $g$ & 13:45--15:33 & 23 & 2823 & 1.11--1.49 & 1.65 & 0.9884 & 0.01184 & 98.2 \\
11 & LRIS & $R_{\rm C}$ & 13:45--15:32 & 16 & 1700 & 1.11--1.48 & 1.29 & 0.9884 & 0.01184 & 98.2 \\
11 & LRIS & $i$ & 14:26--15:30 & 9 & 1222 & 1.11--1.28 & 1.48 & 0.9884 & 0.01184 & 98.1 \\
13 & FORS2 & $g_{\rm HIGH}$ & 07:27--08:12 & 6 & 720 & 1.49--1.86 & 0.97 & 0.9879 & 0.01167 & 98.6 \\
13 & FORS2 & $I_{\rm BESS}$ & 07:31--08:17 & 6 & 1440 & 1.46--1.81 & 0.84 & 0.9879 & 0.01167 & 98.6 \\
14 & FORS2 & $g_{\rm HIGH}$ & 07:50--08:37 & 7 & 1260 & 1.35--1.62 & 0.69 & 0.9876 & 0.01156 & 98.9 \\
14 & FORS2 & $R_{\rm SPECIAL}$ & 07:54--08:44 & 8 & 960 & 1.32--1.59 & 0.64 & 0.9876 & 0.01156 & 98.9 \\
14 & GMOS-N & $i'$ & 13:45--14:54 & 4 & 1664 & 1.22--1.53 & 0.65 & 0.9876 & 0.01153 & 99.0 \\
14 & GMOS-N & $Z$ & 13:52--15:10 & 8 & 3328 & 1.18--1.48 & 0.60 & 0.9876 & 0.01153 & 99.0 \\
15 & FORS2 & $I_{\rm BESS}$ & 07:48--08:52 & 10 & 600 & 1.28--1.62 & 0.73 & 0.9873 & 0.01146 & 99.2 \\
15 & FORS2 & $z_{\rm GUNN}$ & 07:51--08:55 & 10 & 1200 & 1.27--1.60 & 0.72 & 0.9873 & 0.01145 & 99.2 \\
19 & GMOS-N & $r'$ & 14:08--15:21 & 4 & 664 & 1.21--1.48 & 0.65 & 0.9861 & 0.01099 & 101.1 \\
19 & GMOS-N & $i'$ & 14:16--15:12 & 4 & 684 & 1.23--1.43 & 0.61 & 0.9861 & 0.01099 & 101.1 \\
19 & GMOS-N & $Z$ & 14:04--15:25 & 12 & 2820 & 1.20--1.50 & 0.60 & 0.9861 & 0.01099 & 101.1 \\
21 & FORS2 & $I_{\rm BESS}$ & 07:32--08:15 & 10 & 600 & 1.39--1.69 & 0.69 & 0.9856 & 0.01080 & 102.0 \\
21 & FORS2 & $R_{\rm SPECIAL}$ & 07:34--08:17 & 10 & 1200 & 1.38--1.67 & 0.79 & 0.9856 & 0.01080 & 102.0
\enddata
\tablecomments{For the 167 exposures, each row gives the ranges of UTC midpoints and airmasses and the total open-shutter time. Stellar full width at half maximum (FWHM), heliocentric distance $r_{\rm h}$, observatory distance $\Delta$, and solar phase angle $\alpha$ are exposure medians. LRIS $g$ exposures partly overlap the red-arm Cousins $R_{\rm C}$ and SDSS $i$ integrations. GMOS-N $Z$ totals include one trail-crossed exposure per night. Clean stacks total 2912~s on November~14 and 2585~s on November~19. GMOS-N FWHM is measured across the stellar trails.}
\end{deluxetable}
\Needspace{9\baselineskip}
\section{Astrometry used in the orbit fits}\label{app:allastrometry}
\startlongtable
\begin{deluxetable}{l l l l r r c r l r r c}
\renewcommand{\arraystretch}{0.96}
\tabletypesize{\footnotesize}
\setlength{\tabcolsep}{3.1pt}
\tablecaption{Astrometry in the November~22 fit to 71 positions of \qq\label{tab:allastrometry}}
\tablehead{\colhead{UTC date} & \colhead{UTC time} & \colhead{R.A.} & \colhead{Decl.} & \colhead{$\sigma_E$} & \colhead{$\sigma_N$} & \colhead{External fit} & \colhead{Mag.} & \colhead{Band} & \colhead{$(O-C)_E$} & \colhead{$(O-C)_N$} & \colhead{MPC} \\
\colhead{(2026)} & \colhead{(h:m:s)} & \colhead{(h:m:s)} & \colhead{(d:m:s)} & \colhead{($\arcsec$)} & \colhead{($\arcsec$)} & \colhead{$E/N$ ($\arcsec$)} & \colhead{(mag)} & \colhead{} & \colhead{($\arcsec$)} & \colhead{($\arcsec$)} & \colhead{code}}
\startdata
Aug~19 & 22:14:14.352 & 21:38:41.760 & +07:30:41.29 & \nodata & \nodata & 1.323/1.323 & 20.30 & $G$ & $+0.03$ & $-1.28$ & K88 \\
Aug~19 & 22:42:20.189 & 21:38:58.490 & +07:40:28.09 & \nodata & \nodata & 1.323/1.323 & 20.30 & $G$ & $+0.04$ & $-0.74$ & K88 \\
Aug~19 & 22:46:00.768 & 21:39:00.679 & +07:41:44.99 & \nodata & \nodata & 1.323/1.323 & 20.20 & $G$ & $-0.02$ & $-0.27$ & K88 \\
Aug~19 & 22:49:54.912 & 21:39:03.041 & +07:43:06.31 & \nodata & \nodata & 1.323/1.323 & 20.60 & $G$ & $+0.45$ & $+0.00$ & K88 \\
Aug~19 & 23:54:13.190 & 21:39:52.130 & +08:06:47.02 & \nodata & \nodata & 1.000/1.000 & \nodata & \nodata & $+0.50$ & $+0.53$ & L01 \\
Aug~19 & 23:57:08.582 & 21:39:44.141 & +08:06:11.09 & \nodata & \nodata & 1.323/1.323 & 19.60 & $G$ & $+0.54$ & $+0.40$ & K88 \\
Aug~20 & 00:25:36.106 & 21:40:10.990 & +08:17:28.21 & \nodata & \nodata & 1.000/1.000 & 19.60 & $G$ & $+0.63$ & $-0.03$ & L01 \\
Aug~20 & 00:28:52.406 & 21:40:13.020 & +08:18:34.81 & \nodata & \nodata & 1.000/1.000 & 19.70 & $G$ & $+0.94$ & $-0.02$ & L01 \\
Aug~20 & 00:41:39.811 & 21:40:13.121 & +08:21:14.90 & \nodata & \nodata & 1.323/1.323 & 19.80 & $G$ & $-0.10$ & $+0.29$ & K88 \\
Aug~20 & 00:44:37.363 & 21:40:15.170 & +08:22:14.48 & \nodata & \nodata & 1.323/1.323 & 20.00 & $G$ & $+0.60$ & $+0.15$ & K88 \\
Aug~20 & 07:52:52.050 & 21:48:38.873 & +10:58:06.69 & 0.177 & 0.187 & 1.000/1.000 & 20.4 & $G$ & $-0.09$ & $-0.10$ & G37 \\
Aug~20 & 07:57:52.240 & 21:48:41.365 & +10:59:47.90 & 0.239 & 0.259 & 1.000/1.000 & 20.1 & $G$ & $-0.15$ & $-0.18$ & G37 \\
Aug~20 & 22:45:18.480 & 22:06:36.588 & +15:59:01.82 & 0.261 & 0.263 & 1.000/1.001 & 20.0 & $G$ & $-0.35$ & $+0.23$ & Z31 \\
Aug~20 & 22:48:58.820 & 22:06:38.775 & +16:00:19.33 & 0.219 & 0.219 & 1.000/1.001 & 19.8 & $G$ & $-0.25$ & $+0.16$ & Z31 \\
Aug~20 & 22:52:37.410 & 22:06:40.910 & +16:01:36.09 & 0.165 & 0.167 & 1.000/1.001 & 20.3 & $G$ & $-0.36$ & $+0.08$ & Z31 \\
Aug~21 & 00:32:01.000 & 22:06:07.531 & +17:08:36.60 & 0.339 & 0.342 & 1.000/1.000 & 20.5 & $G$ & $-0.45$ & $-0.28$ & M21 \\
Aug~21 & 00:58:09.000 & 22:06:20.567 & +17:16:31.42 & 0.653 & 0.653 & 1.000/1.000 & \nodata & \nodata & $-0.32$ & $-0.73$ & M21 \\
Aug~21 & 02:04:41.810 & 22:08:16.166 & +17:05:42.66 & 0.300 & 0.310 & 1.000/1.000 & 20.3 & $G$ & $-0.01$ & $+0.16$ & Z24 \\
Aug~21 & 02:08:21.290 & 22:08:17.897 & +17:06:51.25 & 0.220 & 0.210 & 1.000/1.000 & 19.9 & $G$ & $-0.29$ & $-0.11$ & Z24 \\
Aug~21 & 02:12:00.150 & 22:08:19.648 & +17:07:59.70 & 0.320 & 0.320 & 1.000/1.000 & 20.5 & $G$ & $-0.34$ & $-0.15$ & Z24 \\
Aug~22 & 01:08:45.400 & 22:33:51.105 & +24:00:02.18 & 0.195 & 0.186 & 1.000/1.000 & \nodata & \nodata & $-0.03$ & $-0.26$ & 204 \\
Aug~22 & 01:29:34.700 & 22:34:04.722 & +24:05:45.63 & 0.213 & 0.201 & 1.000/1.000 & 20.7 & $G$ & $+0.08$ & $-0.28$ & 204 \\
Aug~22 & 01:50:16.200 & 22:34:18.604 & +24:11:21.35 & 0.171 & 0.169 & 1.000/1.000 & \nodata & \nodata & $-0.13$ & $-0.22$ & 204 \\
Aug~22 & 23:19:52.032 & 23:00:00.031 & +30:03:56.30 & \nodata & \nodata & 1.000/1.000 & 20.60 & $G$ & $+0.97$ & $+0.91$ & K88 \\
Aug~22 & 23:23:44.707 & 23:00:02.580 & +30:04:55.88 & \nodata & \nodata & 1.000/1.000 & 20.70 & $G$ & $+0.34$ & $+0.14$ & K88 \\
Aug~22 & 23:26:58.762 & 23:00:04.711 & +30:05:46.10 & \nodata & \nodata & 1.000/1.000 & 20.90 & $G$ & $-0.10$ & $+0.20$ & K88 \\
Aug~23 & 00:46:48.300 & 23:01:20.100 & +30:27:17.75 & 0.235 & 0.235 & 1.000/1.000 & 20.8 & $G$ & $+0.25$ & $+0.11$ & 204 \\
Aug~23 & 01:15:57.500 & 23:01:39.287 & +30:34:23.59 & 0.200 & 0.200 & 1.000/1.000 & \nodata & \nodata & $-0.01$ & $-0.05$ & 204 \\
Aug~23 & 02:14:38.500 & 23:02:01.339 & +30:52:08.40 & 0.740 & 0.750 & 1.000/1.000 & 20.21 & $Si$ & $-0.46$ & $+0.06$ & M57 \\
Aug~23 & 02:33:26.900 & 23:02:14.398 & +30:56:08.95 & 1.120 & 1.120 & 1.000/1.000 & 20.49 & $Si$ & $-0.05$ & $+0.31$ & M57 \\
Aug~23 & 02:52:15.900 & 23:02:27.974 & +31:00:03.71 & 1.110 & 1.110 & 1.000/1.000 & 20.78 & $Si$ & $-0.22$ & $+0.01$ & M57 \\
Sep~5 & 11:41:59.800 & 04:32:40.221 & +50:47:32.87 & 0.500 & 0.500 & 1.000/1.000 & 23.8 & $G$ & $-0.27$ & $+0.03$ & G37 \\
Sep~19 & 11:39:23.760 & 06:34:03.697 & +40:01:28.81 & 0.500 & 0.450 & 1.000/1.000 & 23.9 & $G$ & $-0.02$ & $+0.02$ & G37 \\
Oct~9 & 14:08:38.110 & 07:54:42.327 & +25:51:23.55 & 0.062 & 0.062 & 0.100/0.100 & 24.77 & $G$ & $+0.04$ & $+0.02$ & T12 \\
Nov~7 & 07:51:46.512 & 09:27:31.981 & +04:50:43.74 & 0.187 & 0.187 & \nodata & \nodata & $I$ & $-0.18$ & $-0.02$ & 309$^{\dagger}$ \\
Nov~7 & 07:57:24.007 & 09:27:32.119 & +04:50:32.58 & 0.187 & 0.187 & \nodata & \nodata & $I$ & $-0.20$ & $+0.15$ & 309$^{\dagger}$ \\
Nov~7 & 08:10:43.956 & 09:27:32.391 & +04:50:05.38 & 0.187 & 0.187 & \nodata & \nodata & $I$ & $+0.07$ & $-0.09$ & 309$^{\dagger}$ \\
Nov~11 & 13:56:24.570 & 09:43:38.930 & +00:51:03.45 & 0.184 & 0.252 & \nodata & \nodata & $g$ & $-0.03$ & $-0.02$ & T16$^{\dagger}$ \\
Nov~11 & 14:29:09.904 & 09:43:39.014 & +00:49:47.54 & 0.168 & 0.205 & \nodata & \nodata & $i$ & $-0.14$ & $-0.02$ & T16$^{\dagger}$ \\
Nov~11 & 15:00:11.213 & 09:43:38.612 & +00:48:35.43 & 0.205 & 0.234 & \nodata & \nodata & $R_{\rm C}$ & $-0.09$ & $-0.01$ & T16$^{\dagger}$ \\
Nov~13 & 07:44:43.864 & 09:50:38.625 & $-$00:38:34.01 & 0.158 & 0.138 & \nodata & \nodata & $g$ & $-0.09$ & $-0.11$ & 309$^{\dagger}$ \\
Nov~13 & 08:02:48.627 & 09:50:39.249 & $-$00:39:17.83 & 0.160 & 0.135 & \nodata & \nodata & $g$ & $-0.01$ & $+0.06$ & 309$^{\dagger}$ \\
Nov~13 & 08:11:51.841 & 09:50:39.481 & $-$00:39:39.86 & 0.163 & 0.141 & \nodata & \nodata & $g$ & $-0.02$ & $+0.05$ & 309$^{\dagger}$ \\
Nov~14 & 07:54:12.775 & 09:54:44.825 & $-$01:37:02.01 & 0.048 & 0.055 & \nodata & \nodata & $R$ & $+0.01$ & $-0.06$ & 309$^{\dagger}$ \\
Nov~14 & 08:01:49.685 & 09:54:45.089 & $-$01:37:21.06 & 0.044 & 0.047 & \nodata & \nodata & $R$ & $+0.09$ & $+0.01$ & 309$^{\dagger}$ \\
Nov~14 & 08:09:53.703 & 09:54:45.314 & $-$01:37:41.33 & 0.055 & 0.051 & \nodata & \nodata & $R$ & $-0.06$ & $-0.03$ & 309$^{\dagger}$ \\
Nov~14 & 08:17:38.674 & 09:54:45.502 & $-$01:38:00.71 & 0.050 & 0.045 & \nodata & \nodata & $R$ & $-0.06$ & $+0.02$ & 309$^{\dagger}$ \\
Nov~14 & 08:25:15.695 & 09:54:45.659 & $-$01:38:19.89 & 0.063 & 0.063 & \nodata & \nodata & $R$ & $+0.03$ & $-0.07$ & 309$^{\dagger}$ \\
Nov~14 & 08:33:25.824 & 09:54:45.786 & $-$01:38:40.26 & 0.071 & 0.052 & \nodata & \nodata & $R$ & $+0.05$ & $+0.01$ & 309$^{\dagger}$ \\
Nov~14 & 08:41:22.098 & 09:54:45.866 & $-$01:39:00.15 & 0.053 & 0.071 & \nodata & \nodata & $R$ & $-0.05$ & $-0.01$ & 309$^{\dagger}$ \\
Nov~14 & 08:44:15.260 & 09:54:45.895 & $-$01:39:07.37 & 0.062 & 0.075 & \nodata & \nodata & $R$ & $+0.02$ & $-0.02$ & 309$^{\dagger}$ \\
Nov~14 & 13:44:36.310 & 09:55:45.249 & $-$02:00:52.50 & 0.073 & 0.140 & \nodata & \nodata & $i'$ & $-0.06$ & $+0.04$ & T15$^{\dagger}$ \\
Nov~14 & 14:07:47.710 & 09:55:45.677 & $-$02:01:51.35 & 0.083 & 0.092 & \nodata & \nodata & $i'$ & $-0.07$ & $+0.03$ & T15$^{\dagger}$ \\
Nov~14 & 14:30:59.043 & 09:55:45.796 & $-$02:02:50.01 & 0.080 & 0.084 & \nodata & \nodata & $i'$ & $-0.04$ & $+0.02$ & T15$^{\dagger}$ \\
Nov~14 & 14:54:10.343 & 09:55:45.648 & $-$02:03:48.54 & 0.107 & 0.110 & \nodata & \nodata & $i'$ & $+0.01$ & $-0.04$ & T15$^{\dagger}$ \\
Nov~15 & 07:53:27.134 & 09:58:56.297 & $-$02:36:04.11 & 0.108 & 0.115 & \nodata & \nodata & $I$ & $+0.00$ & $+0.15$ & 309$^{\dagger}$ \\
Nov~15 & 08:27:39.836 & 09:58:57.300 & $-$02:37:32.83 & 0.118 & 0.107 & \nodata & \nodata & $I$ & $+0.05$ & $-0.08$ & 309$^{\dagger}$ \\
Nov~19 & 14:07:42.460 & 10:18:04.378 & $-$07:09:53.13 & 0.098 & 0.077 & \nodata & \nodata & $r'$ & $-0.03$ & $+0.03$ & T15$^{\dagger}$ \\
Nov~19 & 14:37:43.260 & 10:18:05.045 & $-$07:11:21.26 & 0.102 & 0.081 & \nodata & \nodata & $r'$ & $-0.02$ & $+0.01$ & T15$^{\dagger}$ \\
Nov~19 & 14:50:34.076 & 10:18:05.183 & $-$07:11:58.53 & 0.125 & 0.101 & \nodata & \nodata & $r'$ & $-0.03$ & $+0.06$ & T15$^{\dagger}$ \\
Nov~19 & 15:20:32.710 & 10:18:05.220 & $-$07:13:24.67 & 0.125 & 0.106 & \nodata & \nodata & $r'$ & $+0.00$ & $+0.04$ & T15$^{\dagger}$ \\
Nov~21 & 07:34:20.073 & 10:26:46.025 & $-$08:52:37.48 & 0.096 & 0.103 & \nodata & \nodata & $R$ & $+0.01$ & $+0.06$ & 309$^{\dagger}$ \\
Nov~21 & 07:38:56.334 & 10:26:46.405 & $-$08:52:52.05 & 0.137 & 0.143 & \nodata & \nodata & $R$ & $+0.01$ & $+0.12$ & 309$^{\dagger}$ \\
Nov~21 & 07:44:00.488 & 10:26:46.806 & $-$08:53:08.27 & 0.082 & 0.083 & \nodata & \nodata & $R$ & $+0.01$ & $-0.01$ & 309$^{\dagger}$ \\
Nov~21 & 07:48:47.014 & 10:26:47.166 & $-$08:53:23.37 & 0.079 & 0.075 & \nodata & \nodata & $R$ & $+0.01$ & $+0.01$ & 309$^{\dagger}$ \\
Nov~21 & 07:53:23.297 & 10:26:47.493 & $-$08:53:37.91 & 0.106 & 0.104 & \nodata & \nodata & $R$ & $-0.03$ & $+0.02$ & 309$^{\dagger}$ \\
Nov~21 & 07:58:29.324 & 10:26:47.842 & $-$08:53:54.05 & 0.096 & 0.100 & \nodata & \nodata & $R$ & $-0.03$ & $-0.03$ & 309$^{\dagger}$ \\
Nov~21 & 08:03:16.045 & 10:26:48.148 & $-$08:54:09.03 & 0.115 & 0.113 & \nodata & \nodata & $R$ & $-0.11$ & $+0.02$ & 309$^{\dagger}$ \\
Nov~21 & 08:07:53.306 & 10:26:48.450 & $-$08:54:23.60 & 0.097 & 0.080 & \nodata & \nodata & $R$ & $+0.14$ & $-0.03$ & 309$^{\dagger}$ \\
Nov~21 & 08:12:29.052 & 10:26:48.727 & $-$08:54:37.96 & 0.124 & 0.097 & \nodata & \nodata & $R$ & $+0.26$ & $+0.01$ & 309$^{\dagger}$ \\
Nov~21 & 08:17:05.025 & 10:26:48.967 & $-$08:54:52.31 & 0.135 & 0.120 & \nodata & \nodata & $R$ & $+0.02$ & $+0.04$ & 309$^{\dagger}$
\enddata
\tablecomments{Topocentric International Celestial Reference System directions at UTC reception times. Daggers mark 37 team positions in eight MPC acknowledgments (26 VLT/309, three Keck/T16, eight Gemini/T15). The MPC had acknowledged the ten November~21 $R$ positions by the 12:58~UTC cutoff and listed them under the same IDs by 13:42~UTC. For team positions, $\sigma_E$ and $\sigma_N$ are the marginal standard deviations of the adopted covariance, including shared terms (Appendix~\ref{app:orbitweights}). For the three Keck positions, each covariance includes an independent 3.0~s midpoint error projected along the apparent sky motion. For external positions, they are the reported MPC errors where available. The external-fit column gives the adopted errors after station and same-night weighting \citep{Veres2017} and propagation of reported timing errors \citep{Farnocchia2022Timing}. Residuals are observed minus computed for the adopted November~22 radial-acceleration fit to the arc through November~21. The east component includes $\cos\delta$. Magnitudes and bands come from the MPC. For entries without magnitudes, the band gives the observing filter if known. Ellipses mark unavailable or inapplicable entries.}
\end{deluxetable}

\Needspace{18\baselineskip}
\section{Photometry and astrometry}\label{app:photred}
\begingroup
\noindent\begin{minipage}{\linewidth}
\centering
\makeatletter\def\@captype{table}\makeatother
\caption{Symbols used across the photometric and orbital analyses}\label{tab:notation}
\small
\setlength{\tabcolsep}{4pt}
\begin{tabular}{@{}l>{\raggedright\arraybackslash}p{.54\linewidth}ll@{}}
\hline\hline
Symbol & Quantity and units & Section & Equation\\
\hline
\multicolumn{4}{@{}l}{\textit{Photometry and physical properties}}\\*
$C,C_0,m_u^0,m_v^0$ & Target color, calibration-star pivot color, and magnitudes in $u,v$ at that pivot (mag). & \ref{app:fors2cal} & \eqref{eq:qq8_color_solution}\\
$c_u,c_v$ & Stellar color coefficients for filters $u$ and $v$ (mag per mag). & \ref{app:fors2cal} & \eqref{eq:qq8_color_solution}\\
$R_b$ & Solar-relative reflectance in band $b$, normalized at $0.55\,\mu$m. & \ref{app:reflectspec} & \\
$H,p_V$ & Absolute $V$ magnitude and visible geometric albedo. & \ref{sec:density} & \\
$D,\rho,m$ & Spherical diameter (m), bulk density (g~cm$^{-3}$), and mass (kg). & \ref{sec:density} & \\
\multicolumn{4}{@{}l}{\textit{Radiation pressure and capture}}\\*
$A_1$ & Fitted outward radial acceleration at 1~au (au~day$^{-2}$). & \ref{sec:srp} & \\
$C_{\rm R},\mathcal A$ & Radiation-pressure coefficient and effective area-to-mass ratio $\mathcal A=C_{\rm R}A/m$ (m$^2$~kg$^{-1}$), for projected area $A$. & \ref{sec:srp}, \ref{sec:density} & \\
$E_{\rm geo}$ & Geocentric two-body orbital energy per unit mass (au$^2$~day$^{-2}$). & \ref{app:capturemethod} & \eqref{eq:qq8-capture-energy-hill}\\
$r_{\rm geo},R_{\rm H}(t)$ & Earth--asteroid distance and Earth's Hill radius at time $t$ (au). & \ref{app:capturemethod} & \eqref{eq:qq8-capture-energy-hill}\\
\hline
\end{tabular}
\end{minipage}\par\smallskip
\endgroup

\subsection{FORS2 calibration and time matching}\label{app:fors2cal}

ESO's CalSelector service \citep{ESO2011CalSelector} identified the raw exposures for our master biases and flats \citep[see also][]{Bolin2025ThreeIDiscovery}. For the missing November~7 $I_{\rm BESS}$ flat, we used each chip's October~16 master flat from our 2025~SC$_{79}$ observations \citep{Bolin2026SC79}, matched in filter, detector, binning, readout mode, and collimator. We fitted trail-integrated Moffat point-spread functions (PSFs) \citep{Moffat1969,Veres2012Trail} to isolated Pan-STARRS1 Data Release~2 stars \citep{Tonry2012,Flewelling2020} for each exposure's zero point and each filter's nightly color coefficient \citep{Smith2002Photometry} (Table~\ref{tab:photcoef}).

For PS1 color $C=m_u-m_v$, the correction in band $b$ is $c_b(C-C_0)$ relative to the stellar pivot $C_0$. At a common zero point, we linearly interpolated the signed $u$ flux to each bracketed $v$ midpoint and averaged both bands with the same bracket weights. The resulting pivot magnitudes $m_u^0,m_v^0$ give
\begin{equation}
 C=\frac{m_u^0-m_v^0-(c_u-c_v)C_0}{1-c_u+c_v}.
 \label{eq:qq8_color_solution}
\end{equation}
Each draw reused an exposure's flux across brackets. We resampled whole stars, keeping each star's measurements together across exposures and filters, and refitted zero points and slopes \citep{Rubin1981Bootstrap}.

\begin{table}[!htbp]
\centering
\caption{Stellar color coefficients in 2026 November}
\label{tab:photcoef}
\begin{tabular}{lcccc}
\hline\hline
UTC date, instrument & $C$ & $C_0$ (mag) & $c_u$ & $c_v$ \\
\hline
7, FORS2 & $i-z$ & 0.20 & $+0.443$ & $+0.193$ \\
13, FORS2 & $g-i$ & 0.65 & $-0.025$ & $+0.116$ \\
14, FORS2 & $g-r$ & 0.45 & $-0.103$ & $+0.204$ \\
14, GMOS-N & $i-z$ & 0.10 & $+0.285$ & $-0.014$ \\
15, FORS2 & $i-z$ & 0.20 & $+0.440$ & $+0.205$ \\
19, GMOS-N & $r-i$ & 0.20 & $+0.064$ & $\cdots$ \\
19, GMOS-N & $i-z$ & 0.10 & $+0.332$ & $+0.026$ \\
21, FORS2 & $r-i$ & 0.20 & $+0.210$ & $+0.230$ \\
\hline
\end{tabular}
\par\smallskip
\parbox{0.75\linewidth}{\footnotesize\textit{Note.} $C$ and $C_0$ are PS1 AB colors. Coefficients are in mag per mag, in the order of the bands in $C$. GMOS-N zero points refer to the central detector (CCD2). On November~19 the $r$ coefficient uses $r-i$, and both the $i$ and $Z$ coefficients use $i-z$.}
\end{table}

\subsection{FORS2 extraction and recovery tests}\label{app:fors2extract}

We removed the stellar field from each exposure by masking \qq\ in the other same-filter images, aligning them on the stars, and subtracting their scaled median \citep[see also][]{Perley2025Transient}. We fitted signed amplitudes to unmasked pixels using the measured seeing and residual target motion \citep{Stetson1987,Veres2012Trail}, and stacked each filter along the predicted motion \citep{Whidden2019Stack}. On November~7 we fitted a common position offset across both bands. Nine $z$ frames had $I$ brackets, with the last $z$ frame used only in the stack. Of 176 injections, 148 yielded complete tracks. The 74 closest in crowding and mask coverage gave median color biases of $+0.046$~mag for both PSF extractions and $-0.095$~mag for aperture photometry. We weighted direct and field-subtracted PSF fits equally and corrected each draw with paired recovery errors from injections into fixed stellar templates.

\begin{figure*}[!tbp]
\centering
\includegraphics[width=0.84\textwidth]{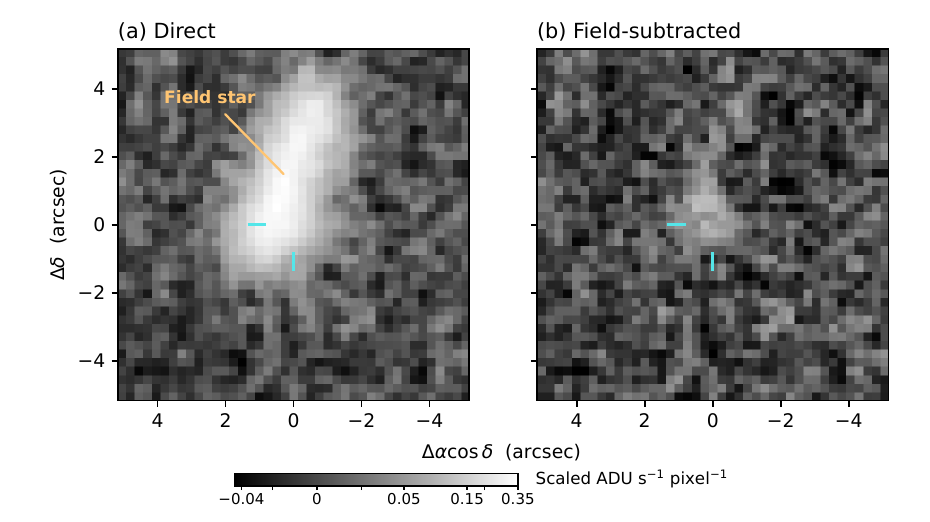}
\caption{FORS2 $g_{\rm HIGH}$ exposure~7 of \qq\ on 2026 November~13 (120~s, midpoint 07:53:59.239 UTC). (a) Direct sky-subtracted image. The orange line marks the PS1 position of a $g_{\rm P1}=21.16$~mag star whose trail crosses the predicted asteroid position. (b) The same image after stationary-field subtraction. Offsets are relative to the Horizons prediction (cyan ticks). Both $10\farcs33$ cutouts have north up, east left, $0\farcs252$ pixels, and the same asinh stretch of pixel fluxes calibrated to a 25~mag zero point. We excluded the two color brackets containing this exposure.}\label{fig:fors2g07}
\end{figure*}

On November~13--15 we fixed the target positions at the ephemeris predictions. A stellar trail raised the direct flux in November~13 $g$ frame~7 by 21 local standard deviations (Figure~\ref{fig:fors2g07}). We excluded both chips, refitted the $g$ calibration and rebuilt the stellar templates. The remaining $g$--$I$--$g$ brackets (1--2--3, 3--4--5, and 9--10--11) gave $g-i=0.680$~mag before injection correction (0.700~mag with the original templates). The adopted color includes calibration error and all 253 tracks (Table~\ref{tab:nightphot}). The 127 crowding-matched tracks gave $0.725^{+0.129}_{-0.170}$~mag. Direct extraction gave $0.926^{+0.163}_{-0.194}$~mag and retained a foreground residual in $I$ frame~4. On November~14, omitting streaked $R$ frame~8 from the templates alone shifted $g-r$ by $-0.002$~mag. On November~15 eight short $I$--$z$--$I$ brackets excluded the long gap and final unbracketed $z$ frame.

On November~21 we interpolated $I$ to the first nine $R$ midpoints and stacked all ten $R$ exposures. For one ADU~s$^{-1}$ at the stellar pivot, zero points spanned 28.058--28.104~mag in $R$ (9--13 standards per exposure) and 27.688--27.721~mag in $I$ (14--18 standards). The median target color lies blueward of the standards' PS1 range, $0.119\leq r-i\leq1.052$~mag. Sampling errors jointly across exposures for 263 complete tracks (285 injected) gave a median color bias of $-0.038$~mag. The 132 crowding-matched tracks gave $(r-i)_{\rm P1}=+0.059$~mag. A trail in $I$ frame~17 raised its direct flux by 2.55 local standard deviations. Omitting this frame from the templates shifted the uncorrected PS1 color from $+0.023$ to $+0.013$~mag. Also dropping its two brackets gave $-0.008$~mag. We retained the field-subtracted measurement.

For the independent November~21 SDSS conversion we used FORS2 responses (pivots 0.655 and 0.795~$\mu$m) and SDSS point-source responses at airmass 1.3 \citep{ESOFORS2Filters,Doi2010SDSS}. In each draw we multiplied a solar-illuminated reflectance template by $(\lambda/0.70\,\mu{\rm m})^s$, choosing $s$ to match the native $R-I$. Photon counts weight the spectrum by $\lambda T_b(\lambda)$ for throughput $T_b$. PS1 solar magnitudes and the draw's stellar coefficients set the native zero points, and 3631~Jy set the SDSS zero points. We applied each band's synthetic offset before differencing the paired magnitudes. We weighted both FORS2 responses equally and assigned half the template weight uniformly across 24 asteroid class means and half uniformly across 21 lunar preparations.

At the median native color, the central 68\% interval for $(r-i)_{\rm SDSS}-(R-I)^0$ is $-0.001$ to $+0.133$~mag. Switching response models shifts the median SDSS color by 0.002~mag. Three choices of wavelength nodes for a two-node linear adjustment give medians of $+0.109$ to $+0.114$~mag. With the primary response, 83 asteroid spectra from \href{https://doi.org/10.1051/0004-6361/202243587}{Mahlke et al.\ (2022)} give $(r-i)_{\rm SDSS}=+0.087$~mag ($-0.007$ to $+0.184$). A featureless power law gives $+0.057$~mag ($-0.026$ to $+0.146$).

\subsection{Keck/LRIS}\label{app:lrisphot}

We fitted signed fluxes in all 48 LRIS images at offsets measured from the filter stacks. We also used LRIS for aperture colors of comet P/2026~N2 \citep{Bolin2026N2} and reflectance spectroscopy of (594913) 'Ayl{\'o}'chaxnim \citep{Bolin2022Ayl}. We weighted $g$ fluxes by their overlap with each red exposure, covering 83\% of the $R_{\rm C}$ and 88\% of the $i$ integration. Direct and field-subtracted fits used 16 and 14 $R_{\rm C}$ images, respectively, and all nine $i$ images. Unweighted $g-R_{\rm C}$ pairs had means and sample standard deviations of $0.86\pm0.35$~mag (direct) and $0.88\pm0.28$~mag (field-subtracted). The nightly colors give equal weight to the two extractions' time-weighted flux sums. Blank tracks (110 direct, 27 field-subtracted) supplied paired background errors. Five stars' residuals against transformed Gaia DR3 magnitudes \citep[][Table~C.2]{Riello2021} gave external uncertainties of 0.101 and 0.096~mag in $g-R_{\rm C}$ and $g-i$ with correlation 0.959.

\subsection{Gemini/GMOS-N}\label{app:gmosphot}

As in our 3I/ATLAS GMOS photometry \citep{Bolin2026ThreeIKeckGemini}, PS1 field stars set the calibration. On November~14 we measured aperture fluxes in four 416~s $i'$ and seven 416~s $Z$ images, excluding $Z$ frame N20261114S0078 because a stellar trail passed $0\farcs24$ from \qq. Cross-trail profiles of stars measured in capsule apertures \citep{Fraser2016TRIPPy} supplied the correction for the asteroid's $0\farcs60$-radius aperture. Evaluating both bands' linear trends at a common time shifted $i'-Z$ by $-0.005$~mag. We combined errors from 129 paired blank tracks with source shot noise and stellar calibration errors.

On November~19 we excluded $Z$ frame N20261119S0164, where a trail crossed \qq, from photometry and stacking. The remaining four $r'$, four $i'$, and eleven $Z$ exposures totaled 664, 684, and 2585~s, respectively. Within $1\farcs94$ of each predicted midpoint we fitted a Moffat profile and sky plane to valid pixels, with one flux per frame and one ratio of target width to cross-trail stellar width per band. At the stellar pivots, $r'-i'=+0.229^{+0.078}_{-0.074}$ and $i'-Z=-0.042^{+0.092}_{-0.097}$~mag. We used Equation~\eqref{eq:qq8_color_solution} to solve for $(i-z)_{\rm P1}$, corrected $i$, then solved the $r$ relation for $(r-i)_{\rm P1}$ (Table~\ref{tab:photcoef}).

We injected sources with shot noise along 91 offset tracks across all 19 images and refitted flux, width, and sky. The 30,000 draws include paired recovery errors, calibration variation from resampling stars, and central-detector reference-star scatter (Table~\ref{tab:nightphot}).

\begin{table}[!htbp]
\centering
\caption{Nightly measurements and combined SDSS-equivalent colors of \qq}
\label{tab:nightphot}
\small
\setlength{\tabcolsep}{4pt}
\begin{tabular}{rlllrr}
\hline\hline
Nov. UTC & Instrument & Photometric system & Quantity & Value & 16th--84th percentiles \\
 & & & & (mag) & (mag) \\
\hline
7  & FORS2 & PS1 AB & $i-z$ & $-0.08$ & $[-0.30,+0.11]$ \\
11 & LRIS & SDSS AB / Cousins Vega & $g-R_{\rm C}$ & $+0.833$ & $[+0.695,+0.965]$ \\
11 & LRIS & SDSS AB & $g-i$ & $+0.942$ & $[+0.666,+1.162]$ \\
11 & LRIS & Cousins Vega / SDSS AB & $R_{\rm C}-i$ & $+0.093$ & $[-0.215,+0.344]$ \\
11 & LRIS & SDSS AB & $r-i$ & $+0.330$ & $[+0.022,+0.581]$ \\
13 & FORS2 & PS1 AB & $g-i$ & $+0.691$ & $[+0.516,+0.865]$ \\
14 & FORS2 & PS1 AB & $g-r$ & $+0.519$ & $[+0.474,+0.570]$ \\
14 & GMOS-N & PS1 AB & $i-z$ & $-0.224$ & $[-0.411,-0.079]$ \\
15 & FORS2 & PS1 AB & $i-z$ & $-0.119$ & $[-0.422,+0.131]$ \\
19 & GMOS-N & PS1 AB & $r$ & $23.997$ & $[23.931,24.045]$ \\
19 & GMOS-N & PS1 AB & $i$ & $23.693$ & $[23.581,23.767]$ \\
19 & GMOS-N & PS1 AB & $z$ & $23.798$ & $[23.729,23.858]$ \\
19 & GMOS-N & PS1 AB & $r-i$ & $+0.303$ & $[+0.204,+0.430]$ \\
19 & GMOS-N & PS1 AB & $i-z$ & $-0.105$ & $[-0.245,+0.027]$ \\
19 & GMOS-N & PS1 AB & $r-z$ & $+0.198$ & $[+0.107,+0.285]$ \\
21 & FORS2 & Native, PS1 pivot & $R^0$ & $23.862$ & $[23.833,23.895]$ \\
21 & FORS2 & Native, PS1 pivot & $I^0$ & $23.798$ & $[23.734,23.888]$ \\
21 & FORS2 & Native, PS1 pivot & $R^0-I^0$ & $+0.058$ & $[-0.026,+0.147]$ \\
21 & FORS2 & PS1 AB & $r$ & $23.834$ & $[23.789,23.877]$ \\
21 & FORS2 & PS1 AB & $i$ & $23.767$ & $[23.718,23.835]$ \\
21 & FORS2 & PS1 AB & $r-i$ & $+0.061$ & $[-0.022,+0.148]$ \\
21 & FORS2 & SDSS AB & $r$ & $23.849$ & $[23.792,23.910]$ \\
21 & FORS2 & SDSS AB & $i$ & $23.743$ & $[23.677,23.806]$ \\
21 & FORS2 & SDSS AB & $r-i$ & $+0.105$ & $[+0.000,+0.217]$ \\
\hline
7--21 & Combined & SDSS AB & $g-r$ & $+0.594$ & $[+0.541,+0.651]$ \\
7--21 & Combined & SDSS AB & $g-i$ & $+0.840$ & $[+0.775,+0.906]$ \\
7--21 & Combined & SDSS AB & $r-i$ & $+0.247$ & $[+0.190,+0.294]$ \\
7--21 & Combined & SDSS AB & $i-z$ & $-0.104$ & $[-0.161,-0.037]$ \\
7--21 & Combined & SDSS AB & $r-z$ & $+0.144$ & $[+0.089,+0.205]$ \\
7--21 & Combined & SDSS AB & $g-z$ & $+0.740$ & $[+0.663,+0.820]$ \\
7--21 & Combined & SDSS AB & $a^\ast$ & $+0.069$ & $[+0.022,+0.119]$ \\
\hline
\end{tabular}
\par\smallskip
\parbox{0.80\linewidth}{\footnotesize\textit{Note.} LRIS $g$ and $i$ are SDSS-equivalent AB magnitudes and $R_{\rm C}$ is Vega-based. Keck colors use all valid shutter overlaps, weighting each $g$ exposure by its overlap with the red exposure (Appendix~\ref{app:lrisphot}). November~19 medians and intervals use joint profile-fit draws, with $r-z=(r-i)+(i-z)$ per draw. The three November~21 calibrations use the same nine time brackets (weighted midpoint 07:52:18 UTC). $R^0,I^0$ are FORS2 magnitudes at the stellar PS1 pivot $(r-i)_{\rm P1}=0.20$~mag.}
\end{table}

\subsection{Passband conversion and joint colors}\label{app:cousins_conversion}\label{app:jointsdss}

The November~14 FORS2 comparison (Section~\ref{sec:colors}) requires converting Keck Cousins $R_{\rm C}$ (Vega) to SDSS $r$ (AB). Integrating solar-illuminated Bus--DeMeo spectra through the Cousins energy and SDSS photon responses \citep{Bessell1990,Doi2010SDSS,BessellMurphy2012} gives $r-R_{\rm C}=0.2153$--$0.2567$~mag across the 24 templates, using CALSPEC \citep{Colina1996} and solar $r-R_{\rm C}=0.220$~mag \citep{Willmer2018}. We adopted the midpoint, $x=(g-r)_{\rm SDSS}=(g-R_{\rm C})-0.2360$. Either endpoint shifts $x$ by 0.021~mag. The stellar transformation of \citet[][Table~6]{Tonry2012} gives
\begin{equation}
 (g-r)_{\rm P1} = x-0.012-0.119x-0.013x^2.
 \label{eq:keck_ps1_gr}
\end{equation}
The 1,520,640 joint Keck draws, including external calibration uncertainty, give $(g-r)_{\rm SDSS}=0.597^{+0.132}_{-0.138}$~mag and $(g-r)_{\rm P1}=0.510^{+0.114}_{-0.119}$~mag.

We compared independent November~21 FORS2 and Keck draws. LRIS $I$ measures SDSS $i$ (near AB). In each paired Keck draw we evaluated $(r-i)_{\rm SDSS}=(g-i)-(g-R_{\rm C})+(r-R_{\rm C})$, sampling the conversion offset uniformly over $[0.2153,0.2567]$~mag, independently of photometric error. Fixing the offset at its endpoints gives median Keck $r-i$ of $+0.308$ and $+0.350$~mag.
We multiplied each trial reflectance by the CALSPEC solar spectrum and integrated through the FORS2 and GMOS-N photon responses \citep{Colina1996,Doi2010SDSS,ESOFORS2Filters,GeminiGMOSFilters}, using the measured stellar zero points and pivots. We scaled each reference spectrum by a positive piecewise-linear function with nodes at 0.475, 0.625, 0.762, and $0.920\,\mu$m, fixing the $0.625\,\mu$m node at unity. We fitted the other three amplitudes to paired November~19 GMOS-N $r'-i'$ and $i'-Z$ draws and independent November~14 FORS2 $g-R$ draws. The remaining colors contributed empirical likelihoods smoothed with a 0.03~mag Gaussian. November~21 entered once through the difference of the paired $R_{\rm SPECIAL}$ and $I_{\rm BESS}$ pivot magnitudes. The alternative throughputs shift the combined SDSS $g-r$, $g-i$, and $i-z$ by less than 0.004~mag each.

For Keck we integrated each spectrum through SDSS $g,i$ (AB) and the Bessell energy response for Cousins $R_{\rm C}$ (Vega), with $(r-R_{\rm C})_\odot=0.220$~mag. We convolved empirical extraction errors with the correlated calibration error for $(g-R_{\rm C},g-i)$. Each extraction method carried half the weight, divided uniformly among its blank tracks (Appendix~\ref{app:lrisphot}). The joint likelihood has means $(0.829,0.915)$~mag, standard deviations $(0.142,0.299)$~mag, and covariance $0.00430$~mag$^2$. Each visit had its own flux normalization. We assigned half the prior weight to each group (asteroid or lunar), uniformly within each. The 350,000 joint SDSS draws sampled each spectrum in proportion to its prior weight times its integrated likelihood.

Substituting physical 2026 LRIS responses and the archived stellar color terms for the calibrated standard passbands shifts the combined $g-r$, $g-i$, and $i-z$ medians by $+0.018$, $+0.001$, and $+0.009$~mag. With these responses, Sq gives $\chi^2=1.39$ and slab 75075,197 gives 2.41.
\subsection{Reflectance and spectral fits}\label{app:reflectspec}

For bands $b,c$, the solar-relative reflectance ratio is $R_c/R_b=10^{0.4[(b-c)-(b-c)_\odot]}$ \citep{Bolin2020CD3}. In each SDSS draw we used $(g-r,g-i,i-z)_\odot=(0.46,0.58,0.03)$~mag \citep{Willmer2018} and normalized at $0.55\,\mu$m by linear interpolation between the $g$ and $r$ reflectances at their pivot wavelengths.

In each SDSS band the template reflectance is the ratio of reflected to incident CALSPEC solar photon flux \citep{Colina1996,Doi2010SDSS}. We converted reflectance ratios between bands to color differences and added the adopted solar colors. Because 25.6\% of the solar-weighted $g$ response lies blueward of the Bus--DeMeo limit ($0.45\,\mu$m), we extrapolated the first measured slope to the filter edge. This linear extension gives the three lowest asteroid $\chi^2$ values for Sq (2.61), K (2.70), and Sv (2.92), in that order. Holding reflectance constant below $0.45\,\mu$m gives Sq (2.23), Sv (2.40), and Sr (2.42). Sq ranks first among the asteroid templates in both cases, and Apollo~17 slab 75075,197 gives the lowest $\chi^2$ overall (1.71).

Let $\boldsymbol{c}$ be the componentwise medians of $(g-r,g-i,i-z)$ and $C$ their covariance across the 350,000 draws. For template colors $\boldsymbol{c}_{\rm mod}$, $\chi^2=(\boldsymbol{c}-\boldsymbol{c}_{\rm mod})^{\mathsf T}C^{-1}(\boldsymbol{c}-\boldsymbol{c}_{\rm mod})$ with three degrees of freedom \citep[][Equation~25]{Hogg2010Fit}. Pairwise tests scored the unsmoothed templates on identical color draws at fixed $C$ (Section~\ref{sec:spectra}). For Figure~\ref{fig:surface} we interpolated class means and dispersion envelopes on a 0.001~nm grid and smoothed them with a $\sigma=12$~nm Gaussian. Lunar and meteorite spectra use a 0.005~nm grid and $\sigma=4$~nm.

\subsection{Lunar samples and albedo}\label{app:lunarpreparations}\label{app:slabalbedo}

The Reflectance Experiment Laboratory (RELAB) spectra \texttt{calr248}--\texttt{calr252} correspond, in order, to Apollo~17 slabs 70215,345, 71055,239, 74275,335, 75075,197, and 78598,7. \citet{Robertson2022Slabs} measured one cut face per specimen at $30^\circ$ incidence and normal emission \citep{Milliken2020RELAB}.

The ordinary-chondrite comparison includes all six H5, seven L5, and five LL5 specimens with grain-size upper limits $\leq150\,\mu$m in \citet[][Table~1]{Dunn2010OC}. Their RELAB spectra cover all four solar-weighted SDSS responses at $30^\circ$ incidence and normal emission. We used them only for the color comparison, with the same covariance as for the lunar samples.
 
\begin{figure*}[!htbp]
\centering
\includegraphics[width=0.83\textwidth]{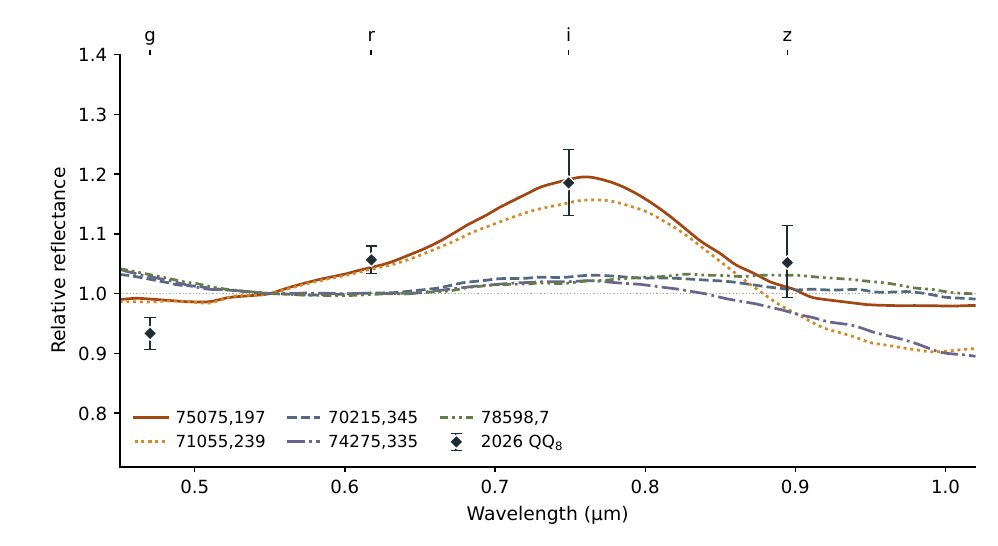}
\caption{Solar-relative reflectance of \qq\ from the combined November~7--21 Keck, Gemini, and VLT colors (diamonds at the SDSS $g,r,i,z$ pivot wavelengths, with 16th--84th percentile bars) and intact cut faces of five Apollo~17 mare basalts \citep{Robertson2022Slabs}. Specimens 75075,197 (solid brown) and 71055,239 (dotted orange) reproduce the measured decline from $i$ to $z$ and give $\chi^2=1.71$ and 2.67, respectively, for the three colors. The display curves use 0.005~nm spacing, Gaussian smoothing with $\sigma=4$~nm, and normalization at $0.55\,\mu$m.}\label{fig:lunarpreparations}
\end{figure*}
The solar-weighted Johnson $V$ reflectance factors of slabs 75075,197 and 71055,239 are $\bar R_V=0.0957$ and 0.1476, respectively \citep{BessellMurphy2012}. For a uniformly covered Lambert sphere, $p_V=2\bar R_V/3$ \citep{MadhusudhanBurrows2012}. A Lommel--Seeliger sphere with constant phase function gives $p_V=0.089$ and 0.138 \citep{Hicks2011Lunar}. The Lommel--Seeliger albedos scale with the unknown ratio of the phase function at $0^\circ$ to its value at $30^\circ$ (Section~\ref{sec:density}).
\subsection{Astrometric reduction and observation selection}\label{app:astrometrymethod}

We propagated Gaia Data Release~3 reference stars from J2016.0 to each exposure midpoint using their proper motions \citep{GaiaDR3,Lindegren2012Core}, including annual parallax for Keck and November~14 FORS2. The stellar plate solution mapped each free target centroid in its ephemeris search region to a topocentric direction (Table~\ref{tab:allastrometry}).

\paragraph{VLT/FORS2.} For the three November~7 positions the median held-out stellar radial rms was $0\farcs045$. The three November~13 positions come from published $g$ frames 05, 09, and 11. On seven of the eight November~14 $R$ images, \qq\ lay outside the area enclosed by the Gaia stars used to calibrate that exposure. We therefore extrapolated the pixel-to-sky transformation to the asteroid. The position uncertainties include residuals of stars near the detector edge withheld from the calibration and shifts in the asteroid centroid after field subtraction. The two November~15 $I$ plates used 25 and 22 stars, with uncertainty floors of $0\farcs053$ and $0\farcs050$.

On November~21 ten $R$ positions (even-numbered frames 2--20) and four $I$ positions (frames 1, 5, 7, and 11) met the single-exposure S/N $\geq5$ and artificial-source recovery criteria. We submitted the $R$ positions to the MPC and used the $I$ centroids as image checks. Fifteen of 20 predictions lay outside the polygons of 20--34 Gaia stars per plate. Dropping the four nearest stars shifted the fitted position of \qq\ by at most $0\farcs056$. The $R$ uncertainties combine 120 injections per frame, held-out stellar residuals, an alternate affine plate, and direct versus field-subtracted centroid shifts. The per-coordinate errors ($0\farcs074$--$0\farcs143$) include a $0\farcs05$ term shared across the night, with a separate 0.2~s timing error along the sky motion (Appendix~\ref{app:orbitweights}).

\paragraph{Keck I/LRIS.} The submitted frames 0187 ($g$), 00199 ($i$), and 00209 ($R_{\rm C}$) used 19, 20, and 35 reference stars, respectively. We fitted trail-averaged Gaussians on the native chips with a common cubic distortion per filter. The midpoint was \texttt{UTSTART} plus half the measured exposure (blue \texttt{ELAPTIME}, red \texttt{XPOSURE}). For the astrometry of \cdthird, \citet[][Section~3.3]{Bolin2020CD3} adopted an LRIS timing uncertainty of $\sim$3~s. We adopted an independent $1\sigma$ midpoint uncertainty of 3.0~s for each \qq\ exposure. The original November~21 solution used 2.0~s for the blue exposure and 0.1~s for each red exposure. We replaced the timing contribution embedded in each error ellipse along the measured apparent motion \citep{Farnocchia2022Timing}.

\paragraph{Gemini North/GMOS-N.} We fitted a Moffat profile and sky plane to each target and used stellar trail midpoints for the tangent-plane solution \citep{Moffat1969,CalabrettaGreisen2002,Shupe2005SIP}. The \href{https://www.gemini.edu/observing/phase-iii-retrieving-reducing-data/reducing-data/timing-information}{Gemini satellite timing correction} shifts the midpoint of \texttt{UTSTART} and \texttt{UTEND} by $-0.207$~s, with $0.020$~s uncertainty shared by all eight submitted positions. On November~14 each 416~s $i'$ plate used 20--25 stars, with quadratic distortion for frame~82. Held-out radial rms was $0\farcs091$--$0\farcs140$, and each position retained its east--north error correlation. For a constant source brightness, curvature of the track shifts the photocenter $0\farcs013$--$0\farcs019$ west of the midpoint prediction.

On November~19 the 20 plates used 12--25 Gaia stars enclosing \qq. Fourfold cross-validation selected 17 linear and three quadratic plates. Held-out radial rms ranged from $0\farcs091$ to $0\farcs179$ (median $0\farcs118$). The position covariance includes held-out residuals, 100 artificial sources per frame, and shifts among five centroid fits. The four submitted $r'$ frames S0153, S0160, S0163, and S0170 enter the orbit at their corrected UTC midpoints.
\section{Orbit determination and numerical integrations}\label{app:orbitmethod}
\setcounter{equation}{0}

\begin{figure}[ht!]
\centering
\includegraphics[width=.91\textwidth,height=.66\textheight,keepaspectratio]{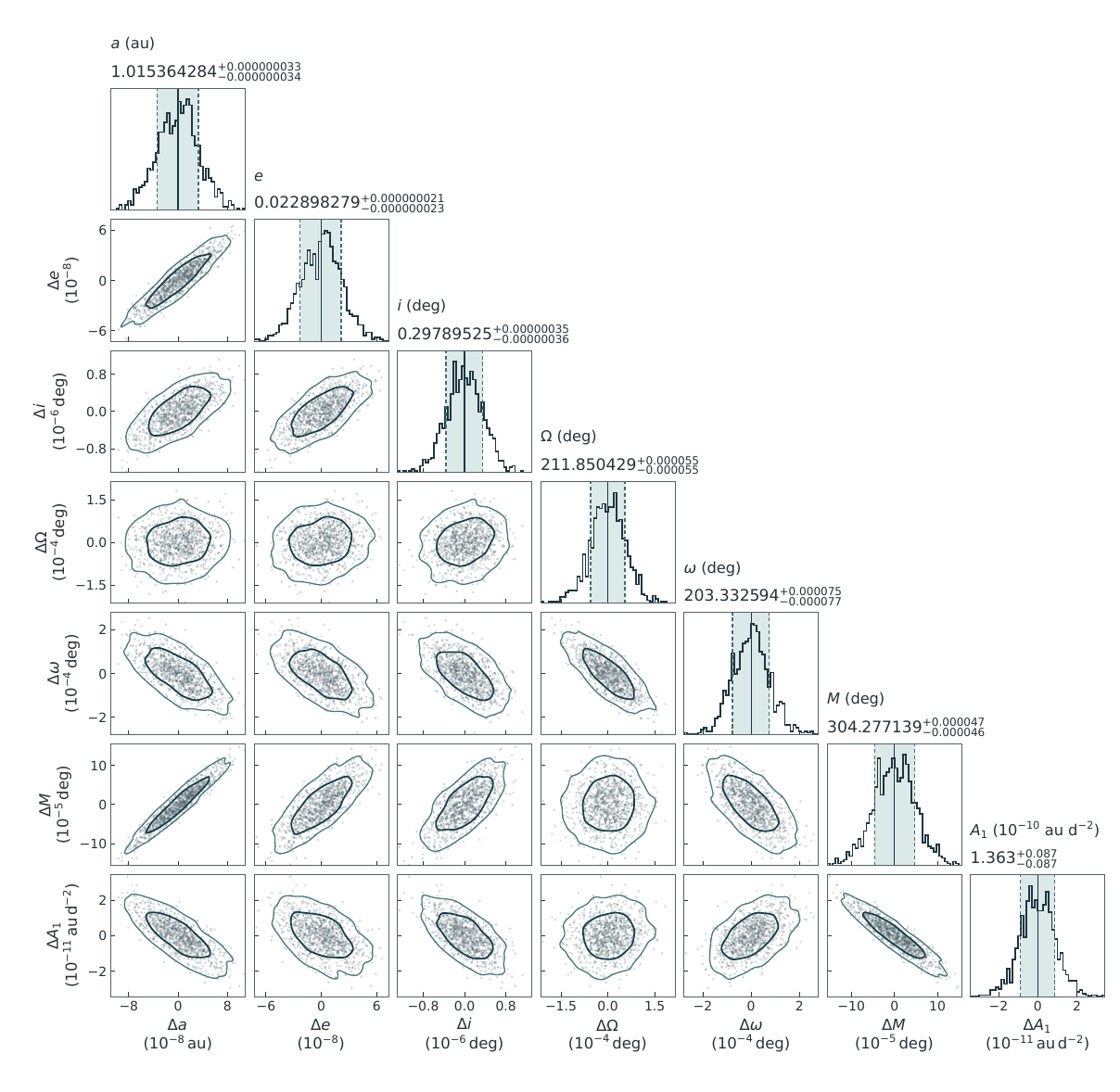}
\caption{Covariance of the November~22 71-position SRP fit with 3.0~s Keck timing uncertainties (Table~\ref{tab:orbitelements}). Points show the 1,000 draws used for the Bura integrations. Axes give heliocentric osculating elements (J2000 ecliptic, JD~2461303.5 TDB) and radial acceleration $A_1$ at 1~au as offsets from their sample means. Contours enclose approximately 68.27\% and 95.45\% of each projected distribution. Diagonal panels mark the mean (solid) and the 15.87th--84.13th percentile interval (shaded, with dashed bounds).}\label{fig:orbitcorner}
\end{figure}

\subsection{Orbit fit and weighting}\label{app:orbitweights}

Using weighted differential corrections \citep{MilaniGronchi2009}, we fitted the six heliocentric Cartesian coordinates at JD~2461303.5 TDB in the International Celestial Reference Frame (ICRF). The seven-parameter fit added a radial acceleration $A_1$ at 1~au varying as inverse-square heliocentric distance \citep{Micheli2012BD}. Starting from the corresponding November~21 solutions, we integrated trial solutions with \texttt{REBOUND}/IAS15 and \texttt{ASSIST} \citep{ReinLiu2012,ReinSpiegel2015,ASSIST}. DE440s supplied the Sun, eight planets, Moon, and Pluto system \citep{Park2021DE440}. We propagated 16 massive asteroids from their epoch states and included solar first post-Newtonian gravity, solar $J_2$, and terrestrial $J_2$--$J_4$. We computed the direction from the observatory at reception to the asteroid at the iterated emission time.

We matched our 37 positions to eight MPC submission receipts. At 12:58 UTC on November~21, 27 of the 63 public positions were ours. The MPC acknowledged the ten new $R$ positions at 12:52 UTC and listed them by 13:42 UTC. Exposure midpoints agreed with receipt epochs (rounded to six decimal days) to within 0.043~s. We weighted east ($\Delta\alpha\cos\delta$) and north residuals with the full measurement covariance. For November~14 FORS2, this included eight measured error ellipses and the stellar plate covariance across exposures. The November~15 and 21 FORS2 ellipses included a $0\farcs05$ nightly term per coordinate, which we propagated between exposures. For other team visits we added a shared $0\farcs05$ per station and night.

We projected the independent 0.2~s FORS2 timing uncertainty along the apparent angular velocity \citep{Farnocchia2022Timing}. Gemini positions retain their measured east--north correlations and a 0.020~s clock uncertainty shared across visits. For each Keck exposure with apparent east--north velocity $\boldsymbol{v}$ in arcsec~s$^{-1}$, we added $(\sigma_{t,{\rm new}}^2-\sigma_{t,{\rm old}}^2)\,\boldsymbol{v}\boldsymbol{v}^{\mathsf T}$ to its original covariance block (Appendix~\ref{app:astrometrymethod}). The updated blocks retain the shared station-night calibration term. The parameter errors are formal uncertainties at the assigned measurement-error scale.

Of the 34 external MPC positions, the Gaia-referenced T12 measurement received $0\farcs10$ per coordinate and the other 33 received $1\arcsec$, including six Tenerife positions at Z24 and Z31 \citep[][Tables~2--4]{Veres2017}. For stations with $N>4$ retained exposures in one local solar night (noon to noon), we multiplied both coordinate uncertainties by $\sqrt{N/4}$. Only K88 met this condition, with seven discovery-night positions weighted at $1\farcs323$ per coordinate (Table~\ref{tab:allastrometry}).

\begin{table}[!htbp]
\centering\footnotesize
\caption{November~22 astrometric fits with and without the November~21 visit}\label{tab:nov21-fit-summary}
\setlength{\tabcolsep}{3pt}
\begin{tabular}{@{}rlrrrr@{}}
\hline\hline
$N$ & Model & $A_1$ & rms & rms$_{21}$ & $\chi^2$/dof \\
\hline
61 & Gravity & $0$ & 1.412 & 0.401 & 231.66/116 \\
61 & Gravity + SRP & $1.370\pm0.097$ & 0.401 & 0.116 & 32.61/115 \\
71 & Gravity & $0$ & 1.345 & 0.259 & 273.57/136 \\
71 & Gravity + SRP & $1.363\pm0.090$ & 0.374 & 0.112 & 43.50/135 \\
\hline
\end{tabular}
\par\smallskip\parbox{0.78\linewidth}{\footnotesize All fits assign independent 3.0~s timing uncertainties to the Keck exposures. $N$ counts fitted positions. $A_1$ is in $10^{-10}$~au~day$^{-2}$ with formal $1\sigma$ errors. The rms and rms$_{21}$ columns give unweighted sky-plane residuals (arcsec) for all fitted positions and the ten November~21 $R$ exposures, respectively. The 61-position fits withhold those ten exposures.}
\end{table}

\begin{table}[htbp]
\centering
\caption{November~22 gravity-only fit to 71 positions spanning 2026 August~19--November~21}\label{tab:gravity-refit}
\begin{tabular}{lrr@{\qquad}lrr}
\hline\hline
Element & Value & Formal $1\sigma$ & Element & Value & Formal $1\sigma$ \\
\hline
Semimajor axis, $a$ (au) & $1.015364656$ & $\pm 0.000000022$ & Ascending node, $\Omega$ (deg) & $211.850350$ & $\pm 0.000059$ \\
Eccentricity, $e$ & $0.022898422$ & $\pm 0.000000019$ & Arg. of perihelion, $\omega$ (deg) & $203.332017$ & $\pm 0.000068$ \\
Inclination, $i$ (deg) & $0.29789846$ & $\pm 0.00000026$ & Mean anomaly, $M$ (deg) & $304.277793$ & $\pm 0.000019$ \\
\hline
\end{tabular}
\par\smallskip
\parbox{0.85\linewidth}{\footnotesize Heliocentric osculating elements in the J2000 ecliptic at JD~2461303.5 TDB (2026 September~20). The uncertainties are marginal standard deviations from the independently fitted six-parameter covariance. The gravity-only control integrations instead start from the matched SRP draws (Appendix~\ref{app:corner}).}
\end{table}

\subsection{Covariance draws}\label{app:corner}

The Bura integrations start from the fitted heliocentric ICRF state $(x,y,z,v_x,v_y,v_z,A_1)$ and its $7\times7$ covariance at JD~2461303.5 TDB (archived as \texttt{submitted\_radial\_A1}). The supplement gives both at full precision along with the ordered positions, $142\times142$ measurement covariance, and east and north residuals. For each of 1,000 Gaussian draws (PCG64, seed 20261111), we applied the Cholesky factor of the fitted correlation matrix to seven independent standard normal deviates, scaled by the parameter standard deviations at the assigned astrometric weights, and added the nominal state \citep{SantanaRos2022XL5}. After conversion to heliocentric osculating elements and unwrapping angles about the nominal orbit, the sample correlations of $A_1$ with $M$ and $a$ were $-0.918$ and $-0.766$ (Figure~\ref{fig:orbitcorner}). Each pair starts from the same sampled Cartesian state, with the sampled $A_1$ in the SRP member and $A_1=0$ in the gravity-only member. Draw IDs 0--999 identify the same pairs across the five bidirectional windows and the forward 10,000~yr integration.

For historical comparison, the November~15 fit used 55 public and two additional FORS2 positions spanning 87.43~days. Its Cartesian and $A_1$ covariance at JD~2461295.5 TDB gave $A_1=(1.3620\pm0.0623)\times10^{-10}$~au~day$^{-2}$. JPL solution~6 used 59 positions spanning 88~days and gave $A_1=(1.3050\pm0.1188)\times10^{-10}$~au~day$^{-2}$, with covariance in $(e,q,T_p,\Omega,\omega,i,A_1)$ at JD~2461303.5 TDB.

\subsection{Capture boundaries}\label{app:capturemethod}

For each clone we selected the continuous interval containing the fit epoch with $E_{\rm geo}<0$ and $r_{\rm geo}<3R_{\rm H}(t)$ \citep{Granvik2012,Fedorets2017}, where
\begin{equation}
 E_{\rm geo}=\frac{v_{\rm geo}^2}{2}-\frac{GM_\oplus}{r_{\rm geo}},\qquad
 R_{\rm H}(t)=r_{\oplus\odot}(t)\left(\frac{M_\oplus}{3M_\odot}\right)^{1/3}.
 \label{eq:qq8-capture-energy-hill}
\end{equation}
Here $r_{\rm geo}$ and $v_{\rm geo}$ are geocentric distance and speed, $r_{\oplus\odot}$ is Earth's heliocentric distance, $G$ is the gravitational constant, and $M_\oplus$ and $M_\odot$ are the Earth and solar masses. For the 2026 event we interpolated the six-hour relative positions and velocities with cubic Hermite polynomials to locate capture and escape. Between crossings we measured each clone's change in longitude and cumulative turning of its path tangent (Table~\ref{tab:captureuncertainty}).

\begin{table}[htbp]
\centering
\caption{2026 capture of matched clones from the November~22 timing refit}\label{tab:captureuncertainty}
\begin{tabular}{lcc}
\hline\hline
Quantity & Gravity only & Gravity and SRP \\
\hline
April 26 entry (TDB) & 08:03:57 & 07:40:22 \\
16th--84th percentile & 08:02:20--08:05:49 & 07:38:49--07:42:12 \\
December 7 predicted exit (TDB) & 06:36:47 & 06:40:54 \\
16th--84th percentile & 06:36:41--06:36:54 & 06:40:44--06:41:04 \\
Duration (days) & 224.93947 & 224.95872 \\
16th--84th percentile & 224.93815--224.94062 & 224.95738--224.95980 \\
Signed inertial longitude (turns) & 1.37222 & 1.37248 \\
Signed rotating longitude (turns) & 0.76378 & 0.76399 \\
3D rotating path curvature (turns) & 1.32907 & 1.32923 \\
\hline
\end{tabular}
\par\smallskip
\parbox{0.85\linewidth}{\footnotesize Medians and 16th--84th percentiles across the 1,000 complete 2026 captures per force law. Signed longitude is the change in unwrapped geocentric ecliptic longitude in fixed J2000 axes or axes rotating with the Sun--Earth line. The three-dimensional curvature measures cumulative turning of the tangent in the rotating frame.}
\end{table}

\subsection{Integration and sampling}\label{app:longintegration}

We integrated the same 1,000 matched pairs over five independent windows of $\pm5$, $\pm10$, $\pm20$, $\pm50$, and $\pm100$ Julian years using \texttt{REBOUND}~4.6.0, IAS15 ($\epsilon=10^{-9}$, initial trial step 0.1~day), and \texttt{ASSIST}~1.2.3 \citep{ReinLiu2012,ReinSpiegel2015,ASSIST}. DE440 and SB441-N16 supplied the Sun, eight planets, Moon, Pluto, and 16 massive asteroids \citep{Park2021DE440}. We tested for capture daily and every six hours when either endpoint lay within five Earth Hill radii. We refined the time of first Earth or Moon contact using surface radii of 6,378.137 and 1,737.4~km. All six SRP clones that struck Earth within the first forward century had gravity-only siblings at the $+100$~yr handoff. SRP clone~405 struck Earth at $-99.00$~yr and survived to $+10{,}000$~yr on its forward branch.

Our Earth-corotating axes rotate about J2000 ecliptic north, with positive $x$ along Earth's heliocentric longitude. For path coverage we used daily positions and three-hour samples within 365 days of the epoch. For near-Earth projections we subtracted Earth's position. We computed the median and 16th--84th percentiles of three-dimensional separation over pairs surviving at each epoch.

At $+100$~yr (JD~2497828.5 TDB) we continued each surviving member from its own Cartesian state with its assigned $A_1$. We initialized the 27 massive bodies from their ephemeris states and integrated their mutual Newtonian gravity with IAS15 through $+10{,}000$~yr, including the solar two-body first post-Newtonian correction. The initial trial step was 0.05~day with $\epsilon=10^{-9}$. The first surface crossing between accepted steps ended a trajectory at the Sun, a planet, the Moon, or Pluto. The 16 perturbing asteroids were point masses. We sampled orbital elements every half year and sibling separations every 25~yr after handoff. Maps accumulate residence time or projected distance per cell, divided by 1,000 launches per model, from daily positions through $+100$~yr and 16-day samples thereafter up to first contact. The supplementary atlas gives the short-window projections, forward residence and path-length maps, and final-decade trajectories.
\subsection{Sensitivity to the starting orbit}\label{app:fit-sensitivity}
For each outcome (Table~\ref{tab:fit-outcomes}) we subtract its gravity indicator (0 or 1) from its SRP indicator within pair $j$ to obtain $d_j$. The difference in percentage points is $100\bar d$, with paired Monte Carlo standard error $100s_d/\sqrt{1{,}000}$ from the sample standard deviation $s_d$. With the original Keck timing weights, the November~21 fit to the same 71 positions gave 807 gravity and 814 SRP survivors at $+10{,}000$~yr ($+0.70\pm1.79$ percentage points). Table~\ref{tab:fit-separation} compares joint survivors at equal elapsed times.
\par\smallskip\noindent
\begin{minipage}{\linewidth}
\centering
\makeatletter\def\@captype{table}\makeatother
\caption{Paired outcomes at $+10{,}000$ yr for the November~22 timing refit and three historical solutions\label{tab:fit-outcomes}}
\footnotesize
\setlength{\tabcolsep}{4pt}
\renewcommand{\arraystretch}{1.08}
\begin{tabular}{llrrrrrrr}
\hline\hline
{Fit} & {Outcome} & {Gravity} & {SRP} & {Both} & {Gravity only} & {SRP only} & {Neither} & {Difference (p.p.)} \\
\hline
Nov.~22 (71) & Survived & 791 & 815 & 642 & 149 & 173 & 36 & $+2.40\pm1.79$ \\
  & Earth & 208 & 185 & 35 & 173 & 150 & 642 & $-2.30\pm1.80$ \\
  & Moon & 1 & 0 & 0 & 1 & 0 & 999 & $-0.10\pm0.10$ \\
  & Any impact & 209 & 185 & 36 & 173 & 149 & 642 & $-2.40\pm1.79$ \\
\hline
Nov.~19 V83 (76) & Survived   & 802 & 813 & 658 & 144 & 155 & 43   & $+1.10\pm1.73$ \\
                 & Earth      & 198 & 187 & 43  & 155 & 144 & 658  & $-1.10\pm1.73$ \\
                 & Moon       &   0 &   0 &  0  &   0 &   0 & 1000 & $0.00\pm0.00$ \\
                 & Any impact & 198 & 187 & 43  & 155 & 144 & 658  & $-1.10\pm1.73$ \\
\hline
Nov.~15 (57) & Survived   & 790 & 826 & 652 & 138 & 174 & 36  & $+3.60\pm1.76$ \\
             & Earth      & 207 & 170 & 32  & 175 & 138 & 655 & $-3.70\pm1.77$ \\
             & Moon       &   3 &   4 &  0  &   3 &   4 & 993 & $+0.10\pm0.26$ \\
             & Any impact & 210 & 174 & 36  & 174 & 138 & 652 & $-3.60\pm1.76$ \\
\hline
JPL 6 (59)   & Survived   & 799 & 802 & 649 & 150 & 153 & 48  & $+0.30\pm1.74$ \\
             & Earth      & 201 & 197 & 48  & 153 & 149 & 650 & $-0.40\pm1.74$ \\
             & Moon       &   0 &   1 &  0  &   0 &   1 & 999 & $+0.10\pm0.10$ \\
             & Any impact & 201 & 198 & 48  & 153 & 150 & 649 & $-0.30\pm1.74$ \\
\hline
\end{tabular}
\par\smallskip\raggedright\textit{Note.}~Each fit uses 1,000 matched pairs. Gravity and SRP give totals by model. The next four columns count pairs in which both, one, or neither member experienced the outcome. The difference is SRP minus gravity in percentage points, with its paired Monte Carlo standard error. Survivors reached the horizon without surface contact. Parentheses give the number of fitted positions. Elapsed time starts at JD~2461303.5 TDB for the November~22 refit, the November~19 (V83) fit, and JPL solution~6, and at JD~2461295.5 TDB for the November~15 fit.
\end{minipage}
\par\smallskip
\par\smallskip\noindent
\begin{minipage}{\linewidth}
\centering
\makeatletter\def\@captype{table}\makeatother
\caption{Three-dimensional separations of jointly surviving gravity and SRP siblings\label{tab:fit-separation}}
\footnotesize
\setlength{\tabcolsep}{4pt}
\renewcommand{\arraystretch}{1.08}
\begin{tabular}{lcccc}
\hline\hline
Fit & $+100$ yr & $+1{,}000$ yr & $+5{,}000$ yr & $+10{,}000$ yr \\
\hline
Nov.~22 (71) & 1.0205 (994) & 1.3569 (873) & 1.3919 (711) & 1.3477 (642) \\
 & [0.4670, 1.6788] & [0.4783, 1.9243] & [0.4723, 1.9269] & [0.5466, 1.9161] \\
Nov.~19 V83 (76) & 1.0901 (992) & 1.4355 (898) & 1.4075 (742) & 1.3970 (658) \\
 & [0.4235, 1.7190] & [0.5786, 1.9177] & [0.5267, 1.9193] & [0.5456, 1.9099] \\
Nov.~15 (57) & 0.9391 (989) & 1.4722 (893) & 1.3989 (727) & 1.3475 (652) \\
 & [0.4020, 1.6214] & [0.5581, 1.9418] & [0.5104, 1.9277] & [0.5162, 1.9274] \\
JPL 6 (59) & 1.2462 (992) & 1.3898 (881) & 1.3729 (726) & 1.4544 (649) \\
 & [0.4329, 1.8433] & [0.5602, 1.9151] & [0.5093, 1.8972] & [0.5470, 1.9276] \\
\hline
\end{tabular}
\par\smallskip\raggedright\textit{Note.}~Separations are in au. Upper rows give the median and the number of jointly surviving pairs (in parentheses). Lower rows give the 16th and 84th percentiles. Times are elapsed years from each fit epoch (Table~\ref{tab:fit-outcomes}). Clone IDs are matched only within a fit.
\end{minipage}
\par\smallskip
\apptocmd{\thebibliography}{\setlength{\itemsep}{-1pt}}{}{}

\end{document}